\documentclass[twocolumn, switch]{article} 

\usepackage{preprint}

\usepackage{enumitem}
\usepackage{amsmath, amsthm, amssymb, amsfonts}
\usepackage{bm}
\usepackage{amsmath}
\usepackage{graphicx}
\usepackage{subfigure}
\usepackage[algoruled,algo2e]{algorithm2e}

\usepackage{setspace}
\usepackage{amssymb}
\usepackage{booktabs}
\usepackage{array}
\usepackage{color}
\usepackage{multirow}
\usepackage{multicol}
\usepackage{threeparttable}
\usepackage{url}
\usepackage[page]{appendix}

\usepackage[numbers,square]{natbib}
\usepackage[utf8]{inputenc}	
\usepackage[T1]{fontenc}	
\usepackage{xcolor}		
\usepackage[colorlinks = true,
            linkcolor = blue,
            urlcolor  = blue,
            citecolor = blue,
            anchorcolor = black]{hyperref}	
\usepackage{booktabs} 		
\usepackage{nicefrac}		
\usepackage{microtype}		
\usepackage{lineno}		
\usepackage{float}			

\usepackage{newfloat}
\DeclareFloatingEnvironment[name={Supplementary Figure}]{suppfigure}
\usepackage{sidecap}
\sidecaptionvpos{figure}{c}

\usepackage{titlesec}
\titlespacing\section{0pt}{12pt plus 3pt minus 3pt}{1pt plus 1pt minus 1pt}
\titlespacing\subsection{0pt}{10pt plus 3pt minus 3pt}{1pt plus 1pt minus 1pt}
\titlespacing\subsubsection{0pt}{8pt plus 3pt minus 3pt}{1pt plus 1pt minus 1pt}

\usepackage{graphics}
\usepackage{hyperref}
\usepackage{color}
\usepackage{multirow}
\usepackage{hhline}
\usepackage{subfigure}
\usepackage{lipsum}

\title{Multi-Auxiliary Augmented Collaborative Variational Auto-encoder for Tag Recommendation}

\usepackage{authblk}

\author[1]{Jing Yi}
\author[1]{Xubin Ren}
\author[1,2,*]{Zhenzhong Chen}
\affil[1]{School of Computer Science, Wuhan University}
\affil[2]{School of Remote Sensing and Information Engineering, Wuhan University}

\begin{document}

\twocolumn[ 
  \begin{@twocolumnfalse} 
  
\maketitle

\begin{abstract}

Recommending appropriate tags to items can facilitate content organization, retrieval, consumption and other applications, where hybrid tag recommender systems have been utilized to integrate collaborative information and content information for better recommendations. In this paper, we propose a multi-auxiliary augmented collaborative variational auto-encoder (MA-CVAE) for tag recommendation,
which couples item collaborative information and item multi-auxiliary information, i.e., content and social graph, by defining a generative process. Specifically, the model learns deep latent embeddings from different item auxiliary information using variational auto-encoders (VAE), which could form a generative distribution over each auxiliary information by introducing a latent variable parameterized by deep neural network. 
Moreover, to recommend tags for new items, item multi-auxiliary latent embeddings are utilized as a surrogate through the item decoder for predicting recommendation probabilities of each tag, where reconstruction losses are added in the training phase to constrict the generation for feedback predictions via different auxiliary embeddings. In addition, an inductive variational graph auto-encoder is designed where new item nodes could be inferred in the test phase, such that item social embeddings could be exploited for new items. Extensive experiments on MovieLens and citeulike datasets demonstrate the effectiveness of our method.

\end{abstract}

\vspace{0.4cm}

  \end{@twocolumnfalse} 
] 

\newcommand\blfootnote[1]{%
\begingroup
\renewcommand\thefootnote{}\footnote{#1}%
\addtocounter{footnote}{-1}%
\endgroup
}

\section{INTRODUCTION}
\label{section1}

{\blfootnote{Corresponding author: Zhenzhong Chen, E-mail:zzchen@ieee.org}}The activity that users annotate items, such as movies and articles, with some key words is called tagging. Online systems like Movielens\footnote{\url{https://movielens.org/}} allow users to tag items. Appropriate tags can facilitate content organization, retrieval, consumption and so on \cite{toc_tag_related,10.1007/978-3-319-22047-5_29,tr_survey,tois_tag,tois_tag1}. Tag recommenders aim to recommend suitable tags for items to help users tag more easily, which also facilitates the spread of the items by creating convenient annotations. Therefore, it is very important to recommend suitable tags for items, especially in the Internet era that contains a great number of items.

Tag recommenders assist the tagging process by utilizing historical interactions between items and tags which are given to the object by online users. Collaborative information contained in item-tag interactions reflects the users' common perception of the object, which has been shown the effectiveness for tag recommendations \cite{hosvd,ptf,pitf}. Among them, Rendle and Schmidt-Thieme \cite{pitf} proposed a pairwise interaction tensor factorization (PITF) model using pairwise interaction tensor factorization and adapted the Bayesian Personalized Ranking (BPR) framework.
However, the sparse feedback of item-tag interactions and the cold-start problem in tag recommendations make systems using only collaborative information hard to work. Content-based tag recommendations \cite{tweet,iTag} have been widely studied. Hassan \textit{et al.} \cite{Bi-GRU+Att} modeled textual content in scientific articles using bidirectional gated recurrent units (bi-GRUs) with word-level and sentence-level attention mechanisms. Tang \textit{et al.} \cite{iTag} proposed a content-based tag recommender system that utilized GRU layers to extract semantic and structural features in text content, where tag correlations and tag-content overlapping were considered. However, hybrid methods which could benefit from both collaborative information and content information have been less explored.

\begin{figure*}[tbp]
\centering
\includegraphics[scale=0.6]{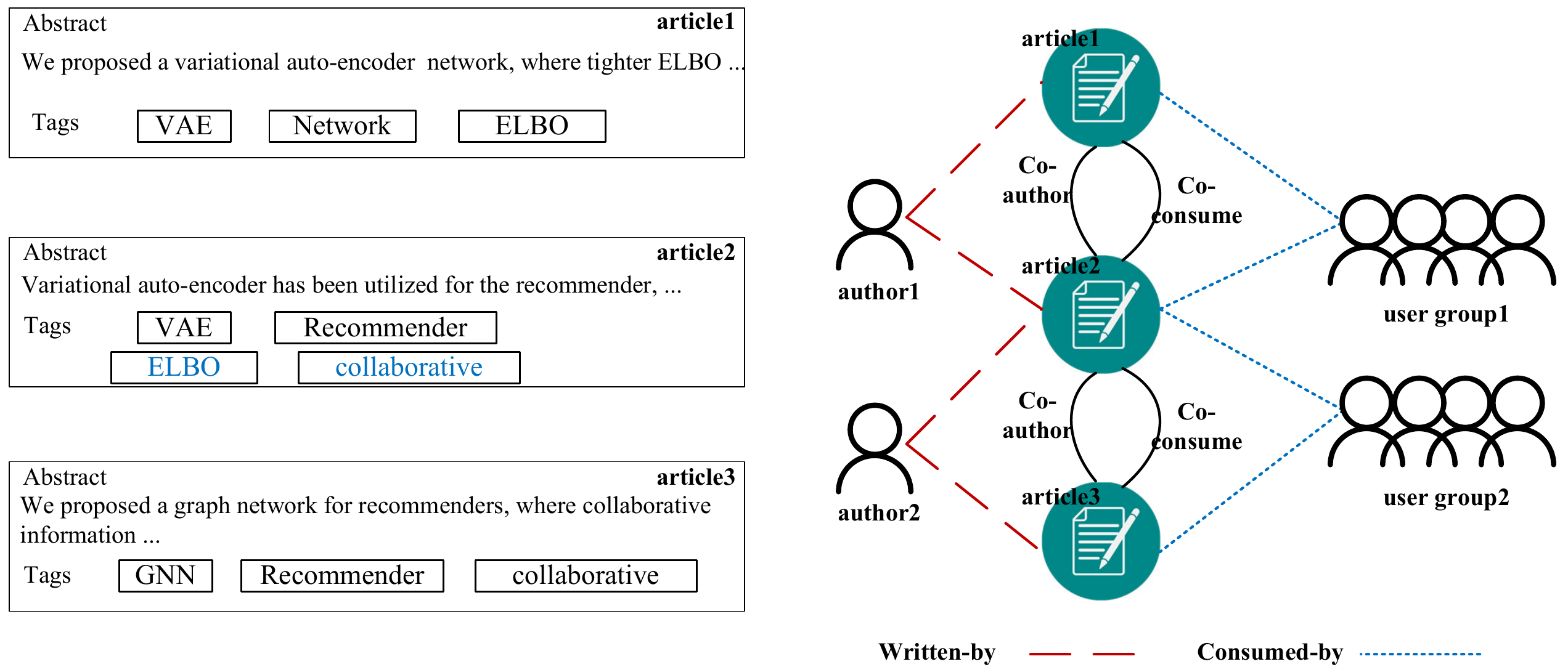}
\caption{On the left is exemplars of item tagging in articles. On the right is the illustration of social interactions between items. Auxiliary information could be obtained by social links among items except for item content information.}
\label{FIG:illu}
\end{figure*}

How to couple item content information and collaborative information is a challenge. Moreover, the item has multiple auxiliary information. Beyond the content information (\textit{e.g.,} textual or visual contents of the item), the social graph information, as illustrated in Fig \ref{FIG:illu}, contains plenty of social-oriented information and fully exploiting them can maximally benefit the tag recommendation process. Collaborative topic regression (CTR) \cite{ctr} used a Bayesian probabilistic model to couple the textual information learned by latent Dirichlet allocation (LDA) and the collaborative information learned by probabilistic matrix factorization (PMF), which has achieved good results. CTR-SR \cite{ctr4tag} extended it by integrating social graph information of items using a linear model with limited representation abilities. Considering the flexibility to integrate different information of Bayesian probabilistic generative process and the powerful feature learning capabilities of deep learning methods, in this paper, we propose to use deep generative models to learn the latent variables of item collaborative and multi-auxiliary information by defining a generative process to couple them. In this way, the collaborative information contained in item-tag interactions and relevant item auxiliary features could be comprehensively extracted with a deep generative model for tag recommendations. Variational auto-encoder (VAE) and variational graph auto-encoder (VGAE) are employed to model the content and social graph information of items respectively, where a generative distribution over each auxiliary information can be obtained by introducing a latent variable parameterized by deep neural network.
Moreover, we substitute the linear PMF model to a multinomial VAE (Mult-VAE) model inspired by \cite{multi-vae} to better capture the item collaborative information from the sparse item-tag ratings. Specifically, we utilize a multinomial likelihood variational auto-encoder which assumes the item-by-tag interactions of each item follow a multinomial distribution such that higher probability mass could be put on the observed interactions. The deep latent variables learned from item multi-auxiliary information are injected into the modeling of item latent embeddings that contains collaborative information, which bridges the hybrid information in a unified framework. Variational inference are utilized for parameter estimation.

Another challenge comes to how to recommend tags for new items which do not have collaborative information during the test phase. Especially for tag recommendations, new items spring up soon. Therefore, extending hybrid recommenders to deal with totally new items is necessary.

To solve the above challenges, we propose a Multi-auxiliary Augmented Collaborative Variational Auto-encoder (MA-CVAE) for tag recommendations. Specifically, we define a probabilistic generative process to couple item collaborative information and multiple auxiliary information into a unified framework, where deep generative models are utilized to model different information. Item auxiliary information consists of various available side information of items, such as content information (\textit{e.g.}, textual or visual content of the item), as well as the social graph information (\textit{e.g.}, co-consumption of users between items, citations between articles, or co-director between movies).
The Product-of-experts (PoE) principle is exploited to integrate item multi-auxiliary Gaussian variables since the mean vector of the product embedding is a weighted sum of the semantic information in content and social graph according to their informative levels for the recommendation. In addition, to solve the item cold-start problem, we extend VGAE to inductive VGAE through sub-graph sampling and neighborhood aggregation so that new item nodes can be inferred during testing. We further add the generative losses of each auxiliary embeddings through the item decoder, where the generation of recommendation probabilities for each tag could be further strengthened via multiple auxiliary information for new items.

The main contributions of our method are as follows:
\begin{itemize}[leftmargin=0pt]
\item By defining a probabilistic generative process, item multiple auxiliary information (content and social graph) are integrated into the modeling of collaborative information, where the closeness of the latent item variable and each auxiliary variable is deduced to be achieved by an mean square error (MSE) loss. Therefore, item auxiliary information could be exploited to make better tag recommendations.
\item Generative losses of different item auxiliary embeddings through the item decoder are added to enhance the prediction ability of interaction feedback. Therefore, the generation of recommendation probabilities for each tag could be well inferred via multi-auxiliary item embeddings for new item scenarios where collaborative information does not exist. Moreover, an inductive variational graph auto-encoder (VGAE) is proposed to solve the problem of adding nodes in the graph by sub-graph sampling and neighborhood aggregation. In this way, item social graph information along with the item content information could be fully exploited to better infer the embeddings of new items for cold-start recommendations.
\item As a Bayesian probabilistic model, the tight coupling between item collaborative information and multiple auxiliary information is realized with deep generative models utilized to better model the multiple data sources. Moreover, extensive experiments have verified the effectiveness of our method in tag recommendations.
\end{itemize}

The rest of the paper is organized as follows. Section \ref{section2} reviews the related work. Section \ref{section3} describes the proposed MA-CVAE model in detail, and Section \ref{section4} presents the experimental evaluations. 
Section \ref{section5} summarizes the paper with conclusions.

\section{RELATED WORK} \label{section2}
In this section, we first review the related work about tag recommendation (TR) from three aspects, \textit{i.e.}, collaborative-based TR that merely relies on the item-tag interaction matrix, content-based TR that utilizes item content information for modeling content filtering, and hybrid methods that merge two information sources. Then, we
summarize the previous work on variational recommender systems where collaborative-based, content-based, and hybrid methods are all included. Specifically, collaborative-based methods utilize variational auto-encoder (VAE) for reconstructing sparse user-aware interactions among items, content-based methods apply the powerful feature extraction ability of VAE for item-side or user-side feature extraction. Hybrid methods seek to utilize both rating information and auxiliary information in a unified framework.

\subsection{Tag Recommendation}
Existing tag recommendation tasks mainly include object-centered and personalized tasks \cite{tr_survey,yuan2019attention}. This paper focuses on the objective-centered tag recommendation without taking users' preferences into account.
From the perspective of exploited data sources, tag recommendation can be roughly categorized into: collaborative-based, content-based, and hybrid methods. Collaborative-based methods utilize the existing tagging history as input. Rendle and Schmidt-Thieme \cite{pitf} presented the factorization model PITF (Pairwise Interaction Tensor Factorization) and adapted the Bayesian Personalized Ranking (BPR) framework. Fang \textit{et al.} \cite{fang2015personalized} exploited the Gaussian radial basis function to increase the model’s capacity which could be considered as a nonlinear extension of Canonical Decomposition. Chen \textit{et al.} \cite{gnn-ptr} integrated the graph neural networks into the pairwise interaction tensor factorization model to better capture the tagging patterns in item-tag interaction graph.

Content-based methods model the semantic and structural content of items to recommend suitable tags. 
Maity \textit{et al.} \cite{maity2019deeptagrec} learned the content representation from question title and body to recommend appropriate question tags on Stack Overflow. Yu \textit{et al.} \cite{osti_10099231} extended labeled latent Dirichlet allocation (LLDA) \cite{llda} by
explicitly specifying several relevant words for a given tag, and allowing to generate the content directly using these words by treating the tags as the supervision information of the corresponding content.
Khezrian \textit{et al.} \cite{khezrian2020tag} used the BERT pre-training technique in tag recommendation task for online Q\&A and open-source communities for the first time. 
Hassan \textit{et al.} \cite{Bi-GRU+Att} adopted deep recurrent neural networks, \textit{i.e.}, bi-GRUs, to encode titles and abstracts of scientific articles into semantic vectors for enhancing the recommendation task. Tang \textit{et al.} \cite{iTag} combined RNN with topical distributions to learn text representations, where the content-tag overlapping and the tag correlation were further considered. Nie \textit{et al.} \cite{tois_tag} constructed a hypergraph by integrating multiple facets, including Question-Answer content analytics, tag-sharing information, as well as user connections, and then selected candidate tags by simultaneously considering
informativeness, stability, and closeness. Nie \textit{et al.} \cite{tois_tag1} further considered the newly-posted question tagging problem by learning the question and topic embeddings from deep neural networks and projecting them into the same
space for a similarity measure.

Hybrid methods integrate collaborative and content information for better recommendations.
Song \textit{et al.} \cite{song2011automatic} introduced a graph-based method, which represented the tagged data
into two bipartite graphs of (document, tag) and (document, word) and found document topics by leveraging graph partitioning algorithms.
Zhang \textit{et al.} \cite{toc} proposed an optimization model to integrate item contents into user interests where different impacts of item features on user preference toward an item have been extracted for item recommendation.
Sun \textit{et al.} \cite{HAM-TR} proposed a hierarchical attention model, where collaborative embeddings and content embeddings were fused through an attention module. Wang \textit{et al.} \cite{ctr} proposed a CTR model to recommend articles to users, which combined a collaborative filtering matrix decomposition algorithm based on hidden factors and a content analysis algorithm based on probabilistic topic models. These two models were unified into a probabilistic generative framework, which could weigh the importance of the content of the article and the collaborative information. For an article that is rarely tagged, it depends more on the content information, and vice versa. Wang \textit{et al.} \cite{ctr4tag} adapted the framework of CTR for tag recommendation problems to seamlessly integrate both item-tag matrix information and item content information, where social networks between items are integrated into the framework for better tag recommendation. Considering the limited representational capabilities of the linear method and topic model to learn interactions and content for the recommendation task, we utilize deep generative models to learn the hidden variables of item collaborative information, content, and social graph. Moreover, by introducing multinomial VAE \cite{multi-vae}, the generative process could be revised to item-based only with an item decoder to predict recommendation probabilities of each tag, where tag latent embeddings are excluded which has more flexibility and conciseness.

\subsection{Variational-based Recommendation}
Variational auto-encoders (VAEs) \cite{kingma2013auto} generate observation $\mathbf{x}$ via the latent variable $\mathbf{z}$.
VAEs construct a variational posterior of the unobserved variable $\mathbf{z}$ given the input $\mathbf{x}$ to approximate the true posterior $p(\mathbf{z} | \mathbf{x})$ due to the nonlinearity of the conditional likelihood where the generative model uses a decoder network to reconstruct $\mathbf{x}$ from ${\mathbf{z}}$.
Specifically, the inference model also uses a neural network as an encoder to learn parameters of approximated posterior distribution 
$q_{\phi}(\mathbf{z} | \mathbf{x})$. By minimizing the Kullback-Leibler divergence between the parametric posterior and the true posterior, the goal of VAEs to maximize the log marginal likelihood $\log p{\mathbf{(x)}}$ deduces to minimize an Evidence Lower BOund (ELBO):
\begin{equation}
    \mathcal{L}\left(\mathbf{x}^{(i)}; \theta, \phi\right) =  \mathbb{E}_{\mathbf{z} \sim q_{\phi}}\left[\log p_{\theta}\left(\mathbf{x}^{(i)} | \mathbf{z}\right)\right]-\mathbb{K L}\left(q_{\phi}\left(\mathbf{z} | \mathbf{x}^{(i)}\right) \| p(\mathbf{z})\right),
    \label{vae}
\end{equation}
where $\mathbf{x}^{(i)}$ is a sample of the observed variable $\mathbf{x}$, $\theta$ and $\phi$ are parameters of the decoder and encoder. The first term on the right-hand side (RHS) of Eq. (\ref{vae}) represents the reconstruction error, which calculates the approximation of observed data and estimated ones to improve the quality of generated data. { The second term on the RHS of Eq. (\ref{vae})} is a regularization term, which penalizes the approximated posterior probability $q_{\phi}\left(\mathbf{z} | \mathbf{x}^{(i)}\right)$ to be far away from prior probability $p(\mathbf{z})$. The reparameterization trick is applied to remove the stochastic sampling from the formation, and thus the gradient could be calculated and back-propagation could be performed. Furthermore, by introducing a hyperparameter $\beta$ before the second term of RHS of Eq. (\ref{vae}), VAE can be extended to beta-VAE \cite{Higgins2017betaVAELB} which controls the balance of reconstruction and regularization terms.

VAEs \cite{kingma2013auto} have been extended for recommendations and other tasks \cite{toc_vae,toc_vae1} due to their advances in modeling high-dimensional data as probabilistic latent variables.
VAE-CF \cite{multi-vae} presented a generative model to fit a user's interactions of items to follow a multinomial likelihood based on beta-VAE. The authors utilized the implicit feedback for items of a user as an input and make predictions based on reconstructed multinomial probability.  MacridVAE \cite{ma2019learning} extended VAE-CF by learning disentangled representations of users' latent preferences at high (\textit{e.g.}, a user's intention towards different categories of items) and low (\textit{e.g.}, a user's preference of the size of clothing) levels. 
Askari \textit{et al.} \cite{askari2021variational} proposed a Joint Variational Auto-encoder (JoVA), which was an
ensemble of two VAEs to jointly learn both user and item
representations to predict user preferences.
{ On the other hand, Li and She \cite{li2017collaborative} exploited VAE  to learn item content variables, and the probabilistic matrix factorization was utilized to model collaborative variables from user-item interactions. By coupling collaborative information and item content information through a Bayesian generative process, better recommendations could be made.}
Chen \textit{et al.} \cite{tois_variational1} proposed a deep generative model, LVSM, to address the item cold-start top-N recommendation problem. The model could capture local aspects of items and measure global item similarity based on deep representations extracted from item features through a variational EM procedure.

VAEs are also utilized for the modeling of hybrid recommendations. Chen and de Rijke \cite{10.1145/3270323.3270326} proposed to simultaneously recover user ratings and side information of items by using a VAE, where user ratings and side information were encoded and decoded collectively through the same inference network and generation network. Due to the heterogeneity of user ratings and side information, the final layer of the generation network followed different distributions. 
Lee \textit{et al.} \cite{aug-vae} proposed to encompass variational auto-encoders through augmenting structures to model the auxiliary information and to model the implicit user feedback, where a conditional VAE \cite{conditional_vae} for modeling the conditional distribution given another modality and a joint multimodal VAE (JMVAE) \cite{suzuki2016joint} for modeling the joint distribution of different modalities by a single latent variable were utilized for integrating multiple item auxiliary features. 
Ma \textit{et al.} \cite{ma2018partial} introduced a partial VAE, which could efficiently handle the missing ratings using amortized partial inference technique without relying on ad-hoc assumptions such as Zero Imputation. The authors further designed a partial inference network for auxiliary distribution by a permutation
invariant set function to encode auxiliary information on user side and item side for better hybrid recommendations.
Wang \textit{et al.} \cite{tois_variational} proposed a personalized online
course recommender system, where the extracted latent representations of the
employees’ competencies from their skill profiles with auto-encoding variational inference based topic modeling and the personal demands of employees were integrated into a unified Bayesian inference view. The graphical model comprehensively combined the conventional latent factor models (LFM)-based collaborative filtering method with auto-encoding variational
inference for topic modeling.

\section{Multi-Auxiliary Augmented Collaborative Variational Auto-encoder}
\label{section3}
\begin{figure*}[tbp]
\centering
\includegraphics[scale=0.55]{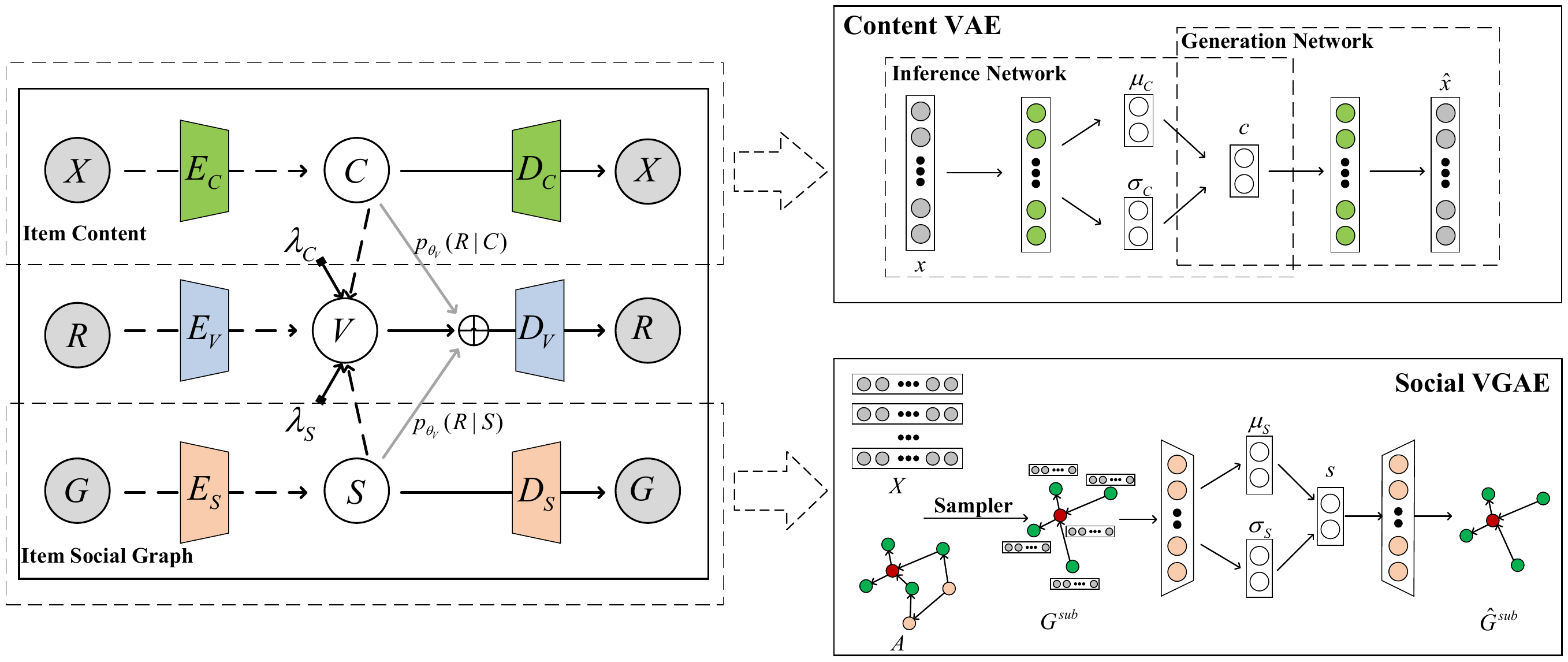}
\caption{On the left is the framework of the proposed MA-CVAE. Item content $\mathbf{x}$ and social graph $\mathbf{G}^{s}$ are different auxiliary information that are augmentations of collaborative information contained in item-tag ratings $\mathbf{r}$. On the right is the zoom-in of the inference
network and generation network of VAE for content information and inductive VGAE for social information. }
\label{FIG:MODEL}
\end{figure*}

In this section, we describe the proposed multiple auxiliary information augmented variational auto-encoder (MA-CVAE) for tag recommendations as shown in Fig. \ref{FIG:MODEL}. MA-CVAE is a generative model where different item auxiliary information, \textit{i.e.,} content and social graph, are generated through item auxiliary variables, and item-tag ratings are generated by item latent variables. Different item auxiliary information is injected into the modeling of item latent variables through adding multi-auxiliary variables and collaborative variables, which bridges hybrid information together into a unified framework. Notations used in this article are summarized in Table \ref{TAB:notation}. Note that we use Capital
non-boldface symbols such as $R$ to denote the corresponding
random vectors of $\mathbf{r}$, $R^m$ is used to denote the random matrix for stacked $\mathbf{r}$. Capital boldface symbols such as $\mathbf{R}$ are used to denote matrices.

\begin{table*}[t]
\centering
\caption{Notations used in our method.}
\setlength{\tabcolsep}{10mm}
\begin{tabular}{ll}
\toprule
  Notation& Description  \\ 
\midrule
$\mathcal{V}$& item set with I items \\
$\mathcal{T}$& tag set with J tags\\
$K$& dimension of latent variables\\
$\mathbf{v}$& latent item variable with the matrix form $\mathbf{V}$\\
$\mathbf{c}$& latent item content variable with the matrix form $\mathbf{C}$ \\
$\mathbf{s}$& latent item social variable with the matrix form $\mathbf{S}$\\
$\mathbf{r}$& the rating vector with the matrix form $\mathbf{R}$ \\ 
$\mathbf{x}$& the content vector with the matrix form $\mathbf{X}$\\
$\mathbf{G}^{s}$& the stacked sub-graph matrix with the full-graph matrix form $\mathbf{G}$\\
$\mathbf{A}$& the adjacency matrix of the full social graph $\mathbf{G}$\\
$V$& random item variable \\
$C$& random item content variable \\
$S$& random item social variable \\
$R$& random rating variable  \\ 
$X$& random content variable\\
$G^{m^s}$&  stacked random social variables for the sub-graph \\
$\theta_v,\theta_c,\theta_s$& trainable generative parameters for item, content and social variables \\
$\phi_v,\phi_c,\phi_s$& trainable inference parameters for item, content and social variables \\
$p_{\theta_v}(R | V)$&  conditional likelihood 
of the ratings, with $\theta_v$ to be trainable weights\\ 
$p_{\theta_c}(X | C)$&  conditional likelihood 
of the content, with $\theta_c$ to be trainable weights\\
$p_{\theta_s}(G^{m^{s}} | S^{m^{s}})$&  conditional likelihoods
of the sub-graph, with $\theta_s$ to be trainable weights \\
$p(V)$& prior of the item random variable \\
$p(C)$& prior of the item content random variable\\
$p(S)$& prior of the item social random variable\\
$q_{\phi_v}(V|R)$& variational posterior of item random variable \\
$q_{\phi_c}(C|X)$& variational posterior of item content random variable\\
$q_{\phi_s}(S^{m^{s}}|G^{m^{s}})$& variational posteriors of item social random variables in a sub-graph\\
\bottomrule
\end{tabular}
\label{TAB:notation}
\end{table*}

\subsection{Problem Statements}
Let $\mathcal{V}$ and $\mathcal{T}$ denote the set of $I$ items and $J$ tags, respectively. Assume we have $\mathbf{C}=\left[\mathbf{c}_{1}, \mathbf{c}_{2}, \cdots, \mathbf{c}_{I}\right]$, where $\mathbf{c}_i$ denotes the content of item $i$. The item social graph $\mathbf{G}$ is a multigraph where multiple edges are distinguished by the attributes of the links. Here, we denote the intrinsic links as the edges between items once after the item is produced such as citations between scientific articles and co-star information between movies. While extrinsic links are denoted as edges between items after user co-consumption of the items such as two items with 5 or more users interacting in common could be linked to an edge. The adjacency matrix is $\mathbf{A} \in \mathcal{R}^{I \times I}$ where we merge these two types of links between items and normalize the edge weights to be \{0, 1\}. The collaborative information is represented by an item-tag tagging matrix $\mathbf{R} = {[r_{ij}]}_{I \times J}$, where $r_{ij} = 1$ means that the tag $j$ is tagged to the item $i$. Given the item $i$, the task of tag recommendation is to find a list of tags $\mathcal{T}_{(i)} \subseteq \mathcal{T}$ that is likely to be annotated to the item $i$.
In this paper, we utilize item content information $\mathbf{X}$, item social graph $\mathbf{G}$, and collaborative information $\textbf{R}$ for hybrid tag recommendations. 

\subsection{Generative Process} 
In our proposed model, we consider item content and social graph to be generated by their latent content variables and social variables through generative networks respectively.
Specifically, to utilize the content information of items, we assign a latent content variable $\mathbf{c}$ for each item. To further employ the social network of items, which contains plenty of contiguous relations between items, we assign a latent social variable $\mathbf{s}$ for each item. 
We draw the latent item content variable and item social variable in a latent low-dimensional space of $K$ dimensions from Gaussian distributions:
\begin{align}
    \mathbf{c} &\sim \mathcal{N}\left(\mathbf{0}, \lambda_C^{-1} \mathbf{I}_{K}\right), \\
    \mathbf{s} &\sim \mathcal{N}\left(\mathbf{0}, \lambda_S^{-1} \mathbf{I}_{K}\right),
\end{align}
where $\lambda_C$ and $\lambda_S$ are precision vectors of item content variable and social variable.
 The content of the item $\mathbf{x}$ is then generated from its latent content variable $\mathbf{c}$ through a generation neural network, \textit{e.g.}, an MLP as with variational auto-encoder \cite{kingma2013auto}.
For the generation of item social graph, sub-graph sampling as in \cite{graphsage} is utilized for inductive learning and mini-batch training. The sub-graph $\mathbf{G}^{s}$ of the item (Here, we focus on the generation of the edges, \textit{i.e.,} the adjacency matrix $\mathbf{A}^{s}$ of the sub-graph) can be generated by latent item social variables $\mathbf{S}^{s}$, which consist of variables of the item and its sampled neighbors, through an inner product decoder as with variational graph auto-encoder \cite{kipf2016variational}. 
Exactly, the content and social graph of the item are generated from its latent content variable $\mathbf{c}$ and social variables $\mathbf{S}^{s}$ through generation neural networks parameterized by $\theta$:
\begin{align}
\mathbf{x} &\sim p_{\theta_c}\left(X | C\right), \\
\mathbf{G}^{s} &\sim p_{\theta_s}\left({G}^{m^s} | {S}^{m^s}\right),
\end{align}
where $\mathbf{G}^{s}$ is a sampled graph from the social graph $\mathbf{G}$ consisting of $l$-hop of neighborhoods of item node, which will be further discussed in the following inductive variational graph auto-encoder section.

To comprehensively utilize the item content and social network information, a product-of-experts (PoE) principle is employed to fuse $\mathbf{c}$ and $\mathbf{s}$ as a multi-auxiliary variable. 
For Gaussian variables, the product is also Gaussian where the new mean becomes ${\mathbf{\mu}_m}  = \left(\mathbf{\mu} _ {c}\lambda_{C} + {\mathbf\mu} _ {s} \lambda_{S}\right) / {\left(  \lambda_{C} + \lambda_{S}\right)}$, and the new variance becomes 
${\mathbf{\lambda}_m}  ={\left( \lambda_{C} \lambda_{S}\right)}/{ \left( \lambda_{C} + \lambda_{S}\right)}$. Since the mean of a Gaussian variable depicts its semantic structure and variance denotes uncertainty theoretically, the mean vector of the item embedding is a weighted sum of the semantic information in content and social graph according to their informative levels for the recommendation.

To fully explore the collaborative information, we explicitly
introduce $\mathbf{v}^{\dagger}$ to embed the item collaborative information for item, and draw it from a Gaussian distribution as:
\begin{equation}
    \mathbf{v}^{\dagger} \sim \mathcal{N}\left(\mathbf{0},  \mathbf{I}_{K}\right).
\end{equation}
Then, we set the latent item variable $\mathbf{v}$ to be composed of both item collaborative and multi-auxiliary latent variables as follows:
\begin{equation}
    \mathbf{v} = \mathbf{v}^{\dagger} + PoE(\mathbf{c}, \mathbf{s}, {\lambda_C}^{-1} \mathbf{I}_{K}, {\lambda_S}^{-1} \mathbf{I}_{K}).
\end{equation}
Given the product of $\mathbf{c}$ and $\mathbf{s}$, then, the latent item variable $\mathbf{v}$ follows the conditional
distribution $\mathcal{N}(PoE(\mathbf{c}, \mathbf{s}, {\lambda_C}^{-1} \mathbf{I}_{K}, {\lambda_S}^{-1} \mathbf{I}_{K}),  \mathbf{I}_{K})$, which is the key to introduce mutual
regularization between $\mathbf{v}$ and $PoE(\mathbf{c}, \mathbf{s}, {\lambda_C}^{-1} \mathbf{I}_{K}, {\lambda_S}^{-1} \mathbf{I}_{K}),  \mathbf{I}_{K})$ in the Maximum A Posterior (MAP) objective.
Moreover, with the PoE principle, we could define the conditional generative process of $\mathbf{v}$ given $\mathbf{c}$ and $\mathbf{s}$ to be:
\begin{equation}
\label{eq:poe}
    p(\mathbf{v} | \mathbf{c}, \mathbf{s}) =  p(\mathbf{v}  |  \mathbf{c}) p(\mathbf{v} | \mathbf{s}).
\end{equation}

\noindent The latent item variable $\mathbf{v}$ is then transformed via a non-linear function, which is parameterized by a neural network, to produce a probability distribution over $J$ tags as in \cite{multi-vae}. The item-by-tag interaction of the item is assumed to be:
\begin{align}
\pi\left(\mathbf{v}\right) &\propto \text{Softmax} \left\{\text{NN}\left(\mathbf{v};\theta_v\right)\right\}, \\
\mathbf{r} &\sim \operatorname{Multinomial}\left(N_{\#}, \pi\left(\mathbf{v}\right)\right),
\end{align}
where $\pi\left(\mathbf{v}\right)$ is computed by a neural network $\text{NN}(\mathbf{v}; \theta_v)$ with the output normalized via a Softmax function. $\theta_v$ is trainable weights of the deep generation network. Given the total number of interactions $N_{\#}$ from the item, $\mathbf{r}$ is sampled from the Multinomial distribution parameterized by $\pi\left(\mathbf{v}\right)$.
The multinomial distribution has been proved to be good at modeling implicit feedback data since the model would assign more probability mass to tags that are more likely to be interacted, which will perform well under the top-$N$ ranking loss that recommender systems are commonly evaluated on.

With the above generative process defined, the joint distribution of observable and hidden variables of the item and its neighbors in the social graph can be formulated as follows:
\begin{equation}
\begin{aligned}
&p_{\theta}(V, C, S^{m^s}, R,  X, G^{m^s})= \\
&p_{\theta_c}(X | C) p(C) p_{\theta_s}(G^{m^s} | S^{m^s}) p(S^{m^s}) p(V | C, S) p_{\theta_v}(R | V),
    \label{eq:joint}
\end{aligned}
\end{equation}
where $\{R,  X, G^{m^s}\}$ is the set of all observed variables, $\{V, C, S^{m^s}\}$ is the set of all latent variables needed to be inferred. The joint distribution of item content variable $p_{\theta_c}(X, C)$ is factorized as $p_{\theta_c}(X | C) p(C)$, with the prior distribution $p(C)$ of latent item content variable to be Gaussian distribution and the stochastic content decoder $p_{\theta_c}(X | C)$ to be parameterized by neural networks. Similarly, the joint distribution of item social variables $p_{\theta_s}(S^{m^s}, G^{m^s})$ is factorized as $p(S^{m^s})$ and the stochastic social decoder $p_{\theta_s}(G^{m^s} | S^{m^s})$. The joint distribution of item variable $p_{\theta_v}(R, V)$ is factorized as the conditional generative distribution $p(V | C, S)$ of latent item variable, and the item decoder $p_{\theta_v}(R | V)$. Moreover, $p_{\theta_s}(G^{m^s} | S^{m^s})$ can be factorized into the product of per node distributions $\Pi_{i} p_{\theta_{s}}\left(G | S\right)$, due to the
assumption of marginal independence among items in the sub-graph.

\subsection{Inference Process}
Since the generative processes of the latent item variable and multiple auxiliary variables are nonlinear (\textit{i.e.}, parameterized as deep neural networks), the posterior distributions for $V$, $C$, and $S^{m^s}$ are intractable. Therefore, we resort to variational inference with an approximated posterior as $q_{\phi} \left(V, C, S^{m^s} | R, X, G^{m^s}\right)$, where we assume the approximated posterior to come from tractable families of distributions (which are also parameterized by deep neural networks) and in those families find the distribution closest to the true posterior measured by the KL-divergence \cite{blei2017variational}.
According to the conditional independence, the joint posterior of all hidden variables can
be factorized into the product of three compact parts with as follows:
\begin{equation}
    q_{\phi}\left(V, C, S^{m^s} | R, X, G^{m^s}\right)= q_{\phi_v}\left(V | R\right) \cdot q_{\phi_c}\left(C | X\right) \cdot q_{\phi_s}\left(S^{m^s} | G^{m^s} \right) ,
\end{equation}
where $q_{\phi_v}\left(V | R\right)$ denotes the variational posterior of item latent variable, $q_{\phi_c}\left(C | X\right)$ and $q_{\phi_s}\left(S^{m^s} | G^{m^s}\right)$ represent the variational posteriors of item content variable and social variables, respectively.  

Previous work \cite{blei2017variational} proves that the minimization of the KL-divergence is equivalent to the maximization of the Evidence Lower BOund (ELBO) \cite{kingma2013auto}, since the marginal likelihood of inputs could be deduced to the KL divergence of the approximate from the true posterior and the ELBO term, where the KL-divergence term is non-negative. The ELBO could be formalized as:
\begin{equation}
\begin{aligned}
\mathcal{L} &= \mathbb{E}_{q_{\phi}} [\log p_{\theta}\left(V,  C, S^{m^s}, R,  X,G^{m^s}\right) 
-\log q_{\phi}\left(V,  C, S^{m^s} | R, X, G^{m^s}\right)] \\
&= \mathbb{E}_{q_{\phi}} [\log p_{\theta_c}(X | C)+\log p_{\theta_s}\left(G^{m^s} | S^{m^s}\right)+\log p\left(V | S\right)+ \\
&\quad \quad \quad  \log p\left(V | C\right) +  \log p_{\theta_v}\left(R | V\right)] \\
&-\mathbb{KL}\left(q_{\phi_c}\left(C | X\right) \| p\left(C\right)\right)-\mathbb{KL}\left(q_{\phi_s}\left(S^{m^s} | G^{m^s}\right) \| p\left(S^{m^s}\right) \right)  \\
&-\mathbb{E}_{q_{\phi_v}} [\log q_{\phi_v}\left(V |R\right)].
\label{eq:elbo}
\end{aligned}
\end{equation}
By substituting the joint distribution of Eq. (\ref{eq:joint}), we can rewrite the ELBO as in Eq. (\ref{eq:elbo}). $\phi$ consists of the parameters of the inference networks where $q_{\phi}$ is an abbreviation for $q_{\phi}\left(V,  C, S^{m^s} | R, X, G^{m^s}\right)$.
As with the CVAE paper \cite{li2017collaborative}, the entropy of $q_{\phi}\left(V | R\right)$ is regarded as a constant and omitted from the ELBO.

\subsection{Maximum A Posterior Estimation}
Maximum A Posterior (MAP) estimation can be performed by considering the
variational distributions of $q_{\phi_c}\left(C | X\right)$, $q_{\phi_s}\left(S^{m^s} | G^{m^s}\right)$ and $q_{\phi_v}\left(V | R\right)$, as well as maximizing the objective with respect to $C$, $S^{m^s}$ and $V$ by an EM-style algorithm using block coordinate ascent. 
Since each variable in our paper is defined as a Gaussian distribution, and the negative logarithmic likelihood of the Gaussian distribution is L2-norm loss, the objective in Eq. (\ref{eq:elbo}) could further be rewrote as:
\begin{equation}
\begin{aligned}
&\mathcal{L}^{\mathrm{MAP}}(V,C,S^{m^s}; \theta, \phi)= \mathbb{E}_{q_{\phi_v}(V | R)} [\log p_{\theta_v}(R | V)]\\
&+\mathbb{E}_{q_{\phi_c}(C | X)} [\log p_{\theta_c}(X | C)]-\mathbb{K L}\left(q_{\phi_c}(C | X) \| p(C)\right)\\
& +\mathbb{E}_{q_{\phi_s}(S^{m^s} | G^{m^s})} [\log p_{\theta_s}(G^{m^s} | S^{m^s})]-\mathbb{K L}\left(q_{\phi_s}(S^{m^s} | G^{m^s}) \| p(S^{m^s})\right)\\
&-\frac{\lambda_{C}}{2}  \mathbb{E}_{q_{\phi}(C,V | X,R)}\left\|V-C\right\|_{F}^{2} \\ &-\frac{\lambda_{S}}{2} \mathbb{E}_{q_{\phi}(S,V | G,R)}\left\|V-S\right\|_{F}^{2} ,
\end{aligned}
\end{equation}
where $\lambda_{C}$ and $\lambda_{S}$ are precision vectors of item content variables and social variables respectively. $F$ is the Frobenius norm and L2-norm are utilized here. $\mathbb{E}_{q_{\phi_v}(V | R)} [\log p_{\theta_v}(R | V)]$, $\mathbb{E}_{q_{\phi_c}(C | X)} [\log p_{\theta_c}(X | C)]$ and $\mathbb{E}_{q_{\phi_s}(S^{m^s} | G^{m^s})} [\log p_{\theta_s}(G^{m^s} | S^{m^s})]$ represent the inference and generation process of item variable, content variable and social variables respectively, in the form of variational auto-encoders. 

We first fix the parameters related to item auxiliary variables $C$, $S$ and optimize the parameters related to item variable $V$ of matrimonial VAE (Mult-VAE). The objective for $V$ thus becomes:
\begin{equation}
\begin{aligned}
\mathcal{L}^{\text{item}}&=\mathbb{E}_{q_{\phi_v}(V | R)} [\log p_{\theta_v}(R | V)]\\
&-\frac{\lambda_{C}}{2}  \mathbb{E}_{q_{\phi_v}(V | R)}\left\|V- \hat{C}\right\|_{F}^{2} -\frac{\lambda_{S}}{2}  \mathbb{E}_{q_{\phi_v}(V | R)}\left\|V-\hat{S}\right\|_{F}^{2},
\label{eq:add_mse}
\end{aligned}
\end{equation}
where $\hat{C}$ equals to the mean vector produced by the item content inference network, $\hat{S}$ equals to the mean vector produced by the item social graph inference network. From Eq. (\ref{eq:add_mse}), we could see that the closeness between item latent embeddings and each latent auxiliary embeddings is achieved by a mean square error (MSE) loss, which could leverage multiple item auxiliary information into the embedding process of collaborative information. In this way, multiple auxiliary information is incorporated to further alleviate the sparsity of the implicit feedback.  Furthermore, to better strengthen the closeness of two components and infer representations for new items using only item auxiliary information, we introduce an additional constraint. Specifically, we add the implementation of the $p_{\theta_v}(R|C)$ and $p_{\theta_v}(R|S)$ to the objective of the item end, using the same decoder as the item decoder $p_{\theta_v}(R|V)$. By adding the additional reconstruction losses, the generation of recommendation probabilities for each tag could be well inferred via item multi-auxiliary embeddings for new items. Therefore, cold-start items that do not contain any collaborative information could be recommended with suitable tags.

Then, we fix the parameters of variable $V$ and item social latent variable $S$, and optimize the item content VAE objective. We isolate the terms related to $C$ and the objective thus becomes:
\begin{equation}
\begin{aligned}
\mathcal{L}^{\text{content}}&=\mathbb{E}_{q_{\phi_c}(C | X)} [\log p_{\theta_c}(X | C)]-\mathbb{K L}\left(q_{\phi_c}(C | X) \| p(C)\right) \\
&-\frac{\lambda_{C}}{2} \mathbb{E}_{q_{\phi_c}(C | X)}\left\|\hat{V}-C\right\|_{F}^{2}.
\label{eq:content}
\end{aligned}
\end{equation}

\noindent Finally, we fix the parameters of item variable $V$ and content variable $C$, the item social inductive VGAE objective can be optimized. The objective for VGAE after isolating the terms related to $S$ becomes:
\begin{equation}
\begin{aligned}
\mathcal{L}^{\text{social}}&=\mathbb{E}_{q_{\phi_s}(S^{m^s} | G^{m^s})} [\log p_{\theta_s}(G^{m^s} | S^{m^s})]-\mathbb{K L}\left(q_{\phi_s}(S^{m^s} | G^{m^s}) \| p(S^{m^s})\right) \\
&-\frac{\lambda_{S}}{2}  \mathbb{E}_{q_{\phi_s}(S^{m^s} | G^{m^s})}\left\|\hat{V^{m^s}}-S^{m^s}\right\|_{F}^{2},
\label{eq:social}
\end{aligned}
\end{equation}
where $\hat{V^{m^s}}$ denotes the stacked mean vector for item variables, which is inferred by the item inference network. For both $\mathcal{L}^{\text{content}}$ and $\mathcal{L}^{\text{social}}$, the objective consists of three parts: 1) the reconstruction part learns a latent Gaussian variable to reconstruct the input; 2) the KL-divergence part assumes a Normal Gaussian as the prior of the latent variable, which penalizes learned latent variable to encode excessive noisy information; 3) the MSE part constricts the closeness of the item variable and each auxiliary variable, which tightly couples the collaborative information and multi-auxiliary information by mutual constraints.

\subsection{Reparameterization Trick}
We utilize reparameterization trick \cite{kingma2013auto} to make the sampling outside the model for amendable gradient-based optimization. For the three Gaussian latent variables $\mathbf{z} \in \{ \mathbf{v}, \mathbf{c}, \mathbf{s}\}$,   we calculate their mean $\mu_z$ and logarithm of standard
deviation $\sigma_z$ via the corresponding encoder network, which could then be scaled by samples drawn from a fixed Gaussian distribution. The transformation is as:
\begin{equation}
    \mathrm{z}\left(\left[\mu_{{z}}, \sigma_{{z}}\right], \epsilon\right)=\mu_{{z}}+\epsilon \odot \sigma_{{z}},
\end{equation}
where $ \epsilon \sim \mathcal{N}\left(\mathbf{0}, \mathbf{I}_{K}\right)$. Through the reparameterization, the random elements are separated, where differentiation is amendable in the backward propagation. According to previous work \cite{kingma2013auto}, we utilize the Monte Carlo gradient estimator to sample for the expectations, which has shown that the variance of the sampling is small and one sample for each data point suffices for convergence.

With the reparameterization trick, we can iteratively optimize the Mult-VAE part, the item content VAE part, and the item social inductive VGAE part for tag recommendation, which forms a tightly coupled hybrid model. That is, collaborative information can guide the learning of item multiple auxiliary feature mapping, and also each auxiliary representation can further improve the performance of interaction-based VAE, especially on items with sparse interactions. In this way, the item content variable and social variable are coupled into the latent item variable with the MSE term. Moreover, the latent item variable that contains collaborative information is further exploited to constrain the updating of the parameters of networks for item content variable and item social variable. Therefore, for sparse datasets which contain less collaborative filtering information, more item content and social network information could be extracted to assist the expression of items, and vice versa. In this way, better recommendations could be made by mutual restraint. Furthermore, $\lambda_S$ and $\lambda_C$ could be viewed as hyper-parameters that balance the item social and auxiliary components.  The training procedure of MA-CVAE is summarized in Algorithm \ref{alg:vae}.

\begin{algorithm2e}[t]
\DontPrintSemicolon  \KwIn{Item-tag interaction matrix $\mathbf{R}$; Item content matrix $\mathbf{X}$; Item social graph $\mathbf{G}=\{\mathbf{A},\mathbf{X}\}$}

Randomly initialize $\theta$, $\phi$.\;
 \While{not converged}{
 \tcp{Update item Mult-VAE.} 
  \ForAll{$v \in \mathcal{V}$}{
  Randomly sample a batch of items with $\mathbf{R}$.\;
    Infer the content mean $\mathbf{C}$ and social mean $\mathbf{S}$.\;
    Sample ${\mathbf{V}}$ via reparametrization trick.\;
    Compute the item loss via Eq. (\ref{eq:add_mse}).\;
  }
  Compute the gradient of the item loss.\;
  Update $\theta_v$, $\phi_v$ by taking stochastic gradient steps.\;
  
  \tcp{Update Content VAE.} 
  \ForAll{$v \in \mathcal{V}$}{
    Randomly sample a batch of items with $\mathbf{X}$.\;
    Infer the item mean $\mathbf{V}$ and social mean $\mathbf{S}$.\;
    Sample ${\mathbf{C}}$ via reparametrization trick.\;
    Compute the content loss via Eq. (\ref{eq:content}).\;
  }
  Compute the gradient of the content loss.\;
  Update $\theta_c$, $\phi_c$ by taking stochastic gradient steps.\;
  
  \tcp{Update Social inductive VGAE.} 
  \ForAll{$v \in \mathcal{V}$}{
    Randomly sample a batch of items for $\mathbf{G}^{s}$.\;
    Infer the item mean $\mathbf{V^s}$ and content mean $\mathbf{C^s}$.\;
    Conduct the neighbor aggregation via Eq. (\ref{eq:graph_}).\;
    Sample ${\mathbf{{S}^s}}$ via reparametrization trick.\;
    Compute the social loss via Eq. (\ref{eq:social}).\;
  }
  Compute the gradient of the social loss.\;
  Update $\theta_s$, $\phi_s$ by taking stochastic gradient steps.\;
}
\Return{$\theta, \phi$}\;
\caption{Training MA-CVAE with SGD.}
\label{alg:vae}
\end{algorithm2e}

\subsection{Inductive Variational Graph Auto-encoder}
To handle the transductive characteristic of variational graph auto-encoder \cite{kipf2016variational} which cannot generalize to unseen nodes by learning embeddings for each node in the graph during training, we propose an inductive VGAE framework inspired by \cite{graphsage}. Specifically, we utilize a graph sampler to sample a sub-graph of the nodes which are to be embedded in the mini-batch, and then an aggregation function is employed to fuse features from the nodes' local neighbors. During training, the parameters of the aggregation function could be well learned through mini-batch gradient descent. Therefore, the embeddings of new nodes which are unseen for the training set could be obtained by leveraging node features of their neighbors using the learned aggregation operation. Meanwhile, the inductive variational graph auto-encoder can be applied for large graphs in recommendation systems such as the user-item bipartite graph through the mini-batch training and inference. 

We exploit the features of items in the item social graph, where textual Term Frequency-Inverse Document Frequency (TF-IDF) attributes of items are utilized as features. We employ an inner-product decoder as the generative process of our inductive VGAE. Inspired by Bayesian personalized ranking (BPR) loss \cite{rendle2012bpr} which is proved to be more suitable for recommendations, the decoder becomes:
\begin{equation}
    {\log p(G^{m^s} | S^{m^s})}=\sum_{\left(v, v^+, v^-\right) \in \mathcal{R}} \log \zeta \left({\mathbf{s}_v}^{ \top} \mathbf{s}_{v^+} -{\mathbf{s}_v}^{ \top} \mathbf{s}_{v^-}\right),
\end{equation}
where $\zeta$ is an activation function to increase nonlinearities and $\mathcal{R}=\left\{\left(v, v^+, v^-\right) |(v, v^+) \in \mathcal{E},\left(v, v^-\right) \notin \mathcal{E}\right\}$ is a set of triplet of three items. $v$ is the target item to be embedded, $v^+$ is the $l$-hop neighbor of item $v$ which are sampled through Random Walk \cite{random-walk} and $v^-$ is the randomly sampled negative item which are not interacted with item $v$.

For the encoder, we utilize a mean aggregation function to leverage features from neighbors, which can be presented as:
\begin{equation}
    \mathbf{s}_{v}^{k} \leftarrow \zeta\left(\mathbf{W} \cdot \operatorname{MEAN}\left(\left\{\mathbf{s}_{v}^{k-1}\right\} \cup\left\{\mathbf{s}_{u}^{k-1}, \forall u \in \mathcal{N}(v)\right\}\right)\right),
\label{eq:graph_}
\end{equation}
where $\mathcal{N}(v)$ is the neighbors of item $v$ in the sampled sub-graph. In this way, the item's previous layer embedding $\mathbf{s}_{v}^{k-1}$ is concatenated with the aggregated neighborhood vector $\mathbf{s}_{u}^{k-1}$ which can be viewed as a skip connection \cite{he2016identity}.
It is worth noting that for new items, links between items contain only intrinsic relations (\textit{e.g.,} citations for articles and co-star for movies) among items without extrinsic links (co-interactions of users for items), since a newly uploaded item is more likely to have no interactions by users. In this way, tag recommendations suggest suitable tags for items and then improve the click-through rate of the items.

\subsection{Prediction}
\subsubsection{For Existing Items}
Let $D$ be the observed data. With the MA-CVAE trained, the weights of the inference networks and generation networks of item implicit feedback, content and social graph, are learned. Then, the prediction for existing items which are existing in the training set becomes:
\begin{equation}
    \mathbb{E}\left[\mathbf{r} | D\right]= p_{{\theta}_v}\left(\mathbf{r} | {\mathbf\mu} _ {v}\right) + p_{{\theta}_v}\left(\mathbf{r} | \frac{ \mathbf{\mu} _ {c}\lambda_{C} + {\mathbf\mu} _ {s} \lambda_{S}}{  \lambda_{C} + \lambda_{S}} \right) ,
\end{equation}
where ${\mathbf\mu} _ {v}$ denotes the mean vector of the latent item variables through the Mult-VAE encoder. ${\left( \mathbf{\mu} _ {c}\lambda_{C} + {\mathbf\mu} _ {s} \lambda_{S}\right)}/{ \left( \lambda_{C} + \lambda_{S}\right)}$ calculates the mean vector of the product of item content variables and social variables.

\subsubsection{For New Items}
For totally new items, the product-of-experts of item content embeddings and social embeddings, which could be easily obtained by the inference network of item content and item social graph, are utilized to make predictions via the item decoder.  The predictions can be represented as:
\begin{equation}
    \mathbb{E}\left[\mathbf{r}| D\right]=  p_{{\theta}_v}\left(\mathbf{r} | \frac{ \mathbf{\mu} _ {c}\lambda_{C} + {\mathbf\mu} _ {s} \lambda_{S}}{  \lambda_{C} + \lambda_{S}}\right).
\end{equation}

\section{EXPERIMENTS}
\label{section4}

To evaluate our proposed MA-CVAE, we conduct extensive experiments to answer the following research questions:
\begin{itemize}[leftmargin=*]
    \item \textbf{RQ1} How does MA-CVAE perform compared with the state-of-the-art methods for tag recommendations? Among them, collaborative-based, content-based, and hybrid methods are all included to make comprehensive comparisons.
    \item \textbf{RQ2} How does MA-CVAE perform under cold-start item scenarios? Comparisons are made among our proposed method and other content-based baselines for new items which are added to the catalog without any interaction. In this situation, collaborative-based models and hybrid methods that are not specifically designed for new items fail to recommend.
    \item \textbf{RQ3} Look inside the proposed MA-CVAE, how is the performance of MA-CVAE affected by the parameters $\lambda_C$ and $\lambda_S$ which influence the balance of item content information and social information? How is the interpretability of MA-CVAE when visualizing some examples?
\end{itemize}

\subsection{Experimental Settings}
\subsubsection{Datasets}
In our experiments, three real-world datasets are utilized to evaluate the effectiveness of our method. 
\begin{itemize}[leftmargin=*]
    \item \textbf{MovieLens 20M} It is a stable baseline dataset\footnote{\url{https://grouplens.org/datasets/movielens/}} for recommender systems, abbreviated as ml-20m, which contains 20 million ratings rated by 138,000 users on 27,000 movies. We focus on the tagging information provided by the dataset and collected the plot of movies as the textual content of items (movies) using the links in IMDB. The crew information is provided by IMDB\footnote{\url{https://www.imdb.com/interfaces/}} and we use the match information provided by ml-20m to correspond a movieId to an imdbId for constructing the item social graph.
    \item \textbf{citeulike-a} This dataset is from CiteULike, which helps users manage academic articles by creating their own collections of articles. On this platform, you can tag and rate your collected references which is the source of our article-tag interactions. We use the collected tags and articles released by \cite{ctr4tag} which contains tagging information, textual information (\textit{i.e.,} title and abstract) and information needed for item social graph (\textit{i.e.,} citations and user-item ratings) in the dataset.
    \item \textbf{citeulike-t} It is an extension of citeulike-a dataset collected by \cite{ctr4tag} which contains more items (articles) and tags. On the other hand, the tagging interactions are more sparse which makes the methods using only collaborative information even harder to perform. 
    
\end{itemize}

\begin{table}[t]
\centering
\caption{Statistics of evaluation datasets after preprocessing.}
\label{TAB:DATA_STATISTIC}
\begin{tabular}{lcccc}
\toprule
Dataset &\#Items &\#Tags &\#Item-Tag &Sparsity   \\ 
\midrule
ml-20m &  1,065 & 992 & 137,748&       13.038\%  \\ 
citeulike-a &   12,734& 11,785& 195,139&  0.130\%   \\ 
citeulike-t&  19,634& 13,162& 220,377&    0.085\%   \\
\bottomrule
\end{tabular}
\end{table}

We preprocess the content of items (\textit{i.e.}, text information) as in \cite{ctr4tag}, where we convert all characters to lowercase, remove the stop words, and conduct lemmatization. Finally, the top 8,000 distinct words are selected as our vocabulary for the citeulike-a dataset. For citeulike-t and ml-20m, the vocabulary sizes are 20,000 and 24,453. We then utilize the TF-IDF of the textual content as our content features for three datasets.

To exploit social information between items, we construct an item social graph using two types of attributes of links (\textit{i.e.,} intrinsic and extrinsic edges). For intrinsic links, we employ citation links existing among items that are available in CiteULike to construct the citation social network (which is a directed graph) for academic articles. While on MovieLens, we exploit co-actor, co-writer, and co-director information as the inherent attributes of movies, where we assign a link between two items if they have one common actor, writer, or director.  On the other hand, the extrinsic links are applied to construct the item co-consumption social graph. Specifically, we employ the co-interacted pattern of users where two items with 4 or more users interacting in common are linked an edge for citeulike, and 50 for ml-20m. We then merge these two networks between items and normalize the weights to be 1 if there exists one edge in any of the two networks. After constructing the item social graph, the number of edges is 368,603, 355,181, and 3,496,719 for citeulike-a, citeulike-t, and ml-20m, respectively.
More specifically, we remove the tags used less than 3 times and the items without any links in the item social graph. The statistics after preprocessing of our three datasets are listed in Table \ref{TAB:DATA_STATISTIC}.

\subsubsection{Evaluation Protocols}
For each dataset, we firstly split out 1000 items for new items which do not exist in the training set. Then, we randomly select tags for each remaining item as training, validation, and test sets with a 6:2:2 ratio. Specifically, items in the training are treated as existing items, and items excluded in the training set are as new items. The tags in the validation and test set are all included in the training set.

We use Recall@$N$, NDCG@$N$, and MRR@$N$ as our evaluation metrics as in \cite{HAM-TR}, where we do not choose Precision@$N$ for the reason that an unobserved tag for an item may be since the tag is not suitable for the item, or that the tag is not considered by users to annotate to the item\cite{ctr}. 
Recall@$N$ calculates the hit ratio of the tested models:
\begin{equation}
    \text{Recall}_{v}@N=\frac{\#\text {Hits }_{v} @ N}{|{y}_{v}|},
\end{equation}
where $|{y}_{v}|$ denotes the number of all interacted tags in test set for item $v$. Hence, $\text{Recall}_{v} @N $ calculates the number of hit tags in top-$N$ recommendation list among Ground Truth of item $v$ and Recall@$N$ averages $\text{Recall}_{v} @N $ for all items in the test set.

Normalized Discounted Cumulative Gain (NDCG) assigns different importance to different ranks which calculates as:
\begin{equation}
    \text{NDCG}_{v} @ N =  {\left(\sum_{i=1}^{N} \frac{2^{r_{i}}-1}{\log _{2}(i+1)}\right)} / {\left(\frac{|{y}_{v}|}{\log2}\right)},
    \label{eq:ndcg}
\end{equation}
where $r_{i}$ represents the relevance degree of the item at position $i$ and it is assigned as binary (\textit{i.e.}, 0 or 1) in implicit recommendation scenario. The numerator of the RHS of Eq. (\ref{eq:ndcg}) calculates the discounted cumulative gain (DCG) score, which would increase according to its ranking position when the item in top-$N$ list hits the Ground Truth $y_v$. Here, the denominator $\frac{|{y}_{v}|}{\log2}$ is to normalize $\text{NDCG}_{v} @ N$ to be in the range [0, 1].

Mean reciprocal rank (MRR) measures where the first correctly predicted tag in the recommendation list appears which calculates as:
\begin{align}
    M R R_v@N=\left\{\begin{array}{l}
\frac{1}{r(T, y_v)}, \text { if } \exists t \in T \text { such that } t \in y_v,\\
0,
\end{array}\right.
\end{align}
where $T$ is a set of recommended top-$N$ tags, $y_v$ is a set of tags that the item $v$ interacted, and $r(T, y_v)$ denotes the position of the first correctly recommended tag in the recommendation list for item $v$. MRR@$N$ calculates the average of $M R R_v@N$ for all items in the test set.

\subsubsection{Baseline Methods}
To evaluate the effectiveness of our model, we compare it with the following state-of-the-art methods for tag recommendations:
\begin{itemize}[leftmargin=*]
    \item \textbf{CF} \cite{mf}  This is a matrix-factorization-based collaborative filtering method that factorizes the training matrix into two low-rank matrices and recovers the original matrix by the inner product of them. It only uses the item-tag matrix information
    \item \textbf{PITF} \cite{pitf} It explicitly models the pairwise interaction between users, items, and tags, which is a strong competitor in the field of personalized tag recommendation.
    \item \textbf{GNN-PTR} \cite{gnn-ptr} It is a graph neural networks boosted personalized tag recommendation model, which integrates the graph neural networks into the pairwise interaction tensor factorization model. 
    \item \textbf{Bi-GRU+Att} \cite{Bi-GRU+Att} This is a content-based tag recommendation method, where deep learning methods are utilized for capturing semantic meanings in the text. Bidirectional gated recurrent units (bi-GRUs) with attention mechanisms are employed to encode text information into semantic vectors.
    \item \textbf{ITAG} \cite{iTag} This content-based tag recommendation method takes tag correlation and content-tag overlapping modeling into consideration beyond capturing textual semantic embeddings using Recurrent Neural Networks.
    \item \textbf{CTR-SR} \cite{ctr4tag} It is a hybrid method that combines the item-tag matrix, item content information, and item social information into a unified framework through a hierarchical Bayesian model.
    \item \textbf{HAM-TR} \cite{HAM-TR} It models two important attentive aspects with a hierarchical attention model, which exploits two levels of attention to effectively aggregate different elements and different information of content information and collaborative information respectively.
\end{itemize}
Here, the first three baselines are collaborative-based methods, where CF and PITF are matrix-based and GNN-PTR is graph-based. Bi-GRU+Att and ITAG are content-based methods. The last two baselines are hybrid methods that exploit both collaborative information and item content information.

\begin{table*}[htbp]
\centering
\caption{Overall performance comparisons between the proposed MA-CVAE model and various baselines.}
\label{TAB:result_with_different_methods}
\setlength{\tabcolsep}{2mm}
\begin{tabular}{l|ccc|ccc|ccc}
\toprule 
 &\multicolumn{3}{c|}{ml-20m } & \multicolumn{3}{c|}{citeulike-a} & \multicolumn{3}{c}{citeulike-t}\\ 
& Recall& NDCG& MRR&  Recall &NDCG&MRR& Recall &NDCG&MRR\\
\midrule
CF  & 0.1504& 0.0861&	0.1714& 	0.0913	&	0.0441&	 0.0831&	0.0812&	0.0365& 0.0576
 \\
PITF& 0.3461& 0.2084& 0.2364 & 	0.1275&	0.0634&	0.1081&	0.1172& 0.0555&	0.0819
 \\
GNN-PTR & 0.3554& 0.2094& 0.2214&	0.1240&	0.0594& 0.0989& 0.1196& 0.0542& 0.0763
 \\ 
Bi-GRU+Att & 0.2521& 0.0199& 0.1632&0.2089&0.0183& 0.1711& 0.2006& 0.0140& 0.1437
 \\
ITAG & 0.2930& 0.0852& 0.0627& 0.1402& 0.0397& 0.0347& 0.1282& 0.0410& 0.0377
 \\
CTR-SR & 0.1280& 0.0601& 0.0965& 0.2052& 0.0847& 0.1189& 0.1167& 0.0479& 0.0537
 \\
HAM-TR & 0.2479& 0.0705& 0.0496&0.2105& 0.0652& 0.0498& 0.1691& 0.0484& 0.0396
\\
 \midrule
$\text{MA-CVAE}_\text{w/o}$& 0.4956& 0.2371& 0.4202& 0.2558& 0.1150& 0.2333&  0.2351& 0.1084& 0.1789
\\ 
$\text{MA-CVAE}_\text{content}$& 0.5539& 0.2721& 0.4813 & 0.3153& 0.1517& 0.3168& 0.3260& 0.1622& 0.2716
\\
$\text{MA-CVAE}_\text{social}$& 0.5555& 0.2730& 0.4829& 0.3266& 0.1575& 0.3273& {0.3363}& \textbf{0.1686}& \textbf{0.2841}     
\\
{MA-CVAE}& \textbf{0.5658}& \textbf{0.2822}& \textbf{0.5000}    & \textbf{0.3277}& \textbf{0.1596}& \textbf{0.3329}& \textbf{0.3369}& {0.1683}&    {0.2818}
\\
\bottomrule
\end{tabular}
\end{table*}

\subsubsection{Parameter Settings}

The validation set is utilized to select the best parameters. Empirically, we set batch\_size and embedding\_size both to be 64, other parameters are found by performing a grid search as follows: $\lambda_C, \lambda_S \in \{0.1, 0.5, 1, 2, 10, 100\}$, $\text{learning\_rate} \in \{0.01, 0.001, 0.0001, 0.0005, 0.00001\}$. The models for collaborative information, content, and social graph are pre-trained in a plain Mult-VAE, VAE, and VGAE manner to first learn initial starting points for the network weights. We set the learning\_rate of pre-trained VAE, pre-trained VGAE, and pre-trained Mult-VAE to be 0.001, empirically. However, the learning\_rates of our MA-CVAE model are carefully turned which controls the balance of the learning speed of collaborative information, content information, and social information. By default, we set our learning\_rate to be 1e-5, 5e-4 for Mult-VAE, and VGAE, respectively for three datasets. And the learning\_rate of content VAE model are set to be 1e-4, 1e-3, and 5e-4 for citeulike-a, citeulike-t, and ml-20m datasets.  The overall architecture for the Mult-VAE is $[J \rightarrow 600 \rightarrow 64 \rightarrow 600 \rightarrow J]$ with $J$ tags. Both
the inference network and the generation network are chosen to
be a two-hidden-layer network architecture ([$64 \rightarrow 64$]) for content VAE network. For VGAE, we set the neighbors to sample for each node in VGAE to be [20, 20] for citeulike-a dataset and [10, 10] for citeulike-t and ml-20m datasets with a two-layer neighborhood aggregation. Moreover, we turn the parameters $\lambda_C$ and $\lambda_S$ which control the balance of content information and social information. We set $\lambda_C$ and $\lambda_S$ both to be 10 by default and further discussion of these two parameters is included in the following section of sensitiveness to parameters.

\subsection{Overall Comparison}
\begin{table*}[htbp]
\centering
\caption{Performance comparisons between the proposed MA-CVAE model and other baselines on cold-start items.}
\label{TAB:result_with_new_items}
\setlength{\tabcolsep}{2mm}
\begin{tabular}{l|ccc|ccc|ccc}
\toprule
 &\multicolumn{3}{c|}{ml-20m } & \multicolumn{3}{c|}{citeulike-a} & \multicolumn{3}{c}{citeulike-t}\\ 
& Recall &NDCG&MRR&  Recall &NDCG&MRR& Recall &NDCG&MRR\\
\midrule
Bi-GRU+Att& 0.2023& 0.0660& 0.2425 & 0.2113& 0.0922& 0.2885& 0.1859& 0.0504& 0.2334
 \\
ITAG & 0.2930& 0.0854& 0.0601& 0.1259& 0.0363& 0.0324& 0.1284& 0.0407& 0.0395
 \\
 \midrule
$\text{MA-CVAE}_\text{content}$ & 0.2647& 0.1167&   0.5940& 0.2605& 0.1176& 0.7566& 0.2640& 0.1233& 0.5317      
\\
$\text{MA-CVAE}_\text{social}$& 0.2840& 0.1247&  0.6029 & 0.2734& 0.1228& 0.7678& 0.2952& 0.1402& 0.5829     
\\
{MA-CVAE}& \textbf{0.2983}& \textbf{0.1313}& \textbf{0.6153}& \textbf{0.2862}& \textbf{0.1291}& \textbf{0.7971}& \textbf{0.3085}& \textbf{0.1477}& \textbf{0.6077}
\\
\bottomrule
\end{tabular}
\end{table*}

The overall performance of our proposed MA-CVAE and the state-of-the-art baselines for tag recommendations are summarized in Table \ref{TAB:result_with_different_methods}, where the upper part is various baselines and the bottom part is our proposed method with ablations. $\text{MA-CVAE}_\text{w/o}$ only conduct Mult-VAE on collaborative information, $\text{MA-CVAE}_\text{content}$ and $\text{MA-CVAE}_\text{social}$ perform with augmented item content information and social information, respectively. Each metric is averaged across all test users and we report top-20 evaluations.

First of all, by comparing collaborative-based methods (\textit{i.e.,} CF, PITF, and GNN-PTR) and content-based methods (\textit{i.e.,} Bi-GRU-Att and ITAG), we can see that content-based methods perform better on sparse datasets like citeulike. However, on more dense datasets like ml-20m, collaborative-based methods perform much better, which demonstrates the effectiveness of collaborative filtering by using co-occurrence information of collaborative-based methods. For the hybrid method CTR-SR, it achieves satisfactory results on the citeulike-a dataset, while on the other two datasets whose content features are high-dimensional and sparse, it cannot perform well. This is because of the weakness of LDA in modeling high-dimensional and sparse features. On the other hand, the other hybrid method HAM-TR, which fuses collaborative and content information with weighted average using latent factor methods for modeling both information, achieves somewhat satisfactory results. 

However, our proposed MA-CVAE significantly outperforms all baselines. Compared to collaborative-based methods, $\text{MA-CVAE}_\text{w/o}$ with only collaborative information outperforms the strong GNN-PTR by a large margin mainly due to the modeling of collaborative information via a deep generative model and applying a multinomial likelihood for the implicit interaction data which has been proved to be good for top-$N$ recommendations. For content-based methods, $\text{MA-CVAE}_\text{content}$ performs better than them due to the effective modeling of collaborative information and content information. $\text{MA-CVAE}_\text{content}$ also outperforms the hybrid method HAM-TR by modeling content with deep generative models and tight coupling of item content information and collaborative information. CTR-SR, which though makes use of item content and social graph information, applies linear methods to model the item auxiliary information. Therefore, it achieves inferior performance compared to our MA-CVAE method. 
We further attribute the performance improvements to the following two reasons: 1) By defining a probabilistic generative process, the tight coupling of collaborative information and item auxiliary information is well achieved by MSE losses between the item latent embeddings and multiple auxiliary embeddings. In this way, the two types of information could be maximally utilized in a tuning way. 2) The modeling of collaborative information and multi-auxiliary information are conducted with deep generative models, which have good representational abilities on sparse data. Moreover, the variational-based methods could be robust to noises.

As an ablation study of our method, $\text{MA-CVAE}_\text{w/o}$ with only collaborative information achieves inferior results compared to $\text{MA-CVAE}_\text{content}$ and $\text{MA-CVAE}_\text{social}$ with item content information and social information, respectively. It demonstrates that different item auxiliary information all play vital roles in tag recommendations, especially the social information on citeulike-t dataset. Moreover, by defining a generative process, we fuse multiple auxiliary information via a product-of-experts module and tightly couple the collaborative information and multi-auxiliary information, which achieves the best performance.

\subsection{Recommendation for Cold-start Items}

We explore the effectiveness of our proposed method when recommending tags to totally new items, which do not exist in the training set. Collaborative-based (\textit{i.e.,} CF, PITF, and GNN-PTR) and hybrid methods (\textit{i.e.,} CTR-SR and HAM-TR) in our baselines are dependent on interactions when making predictions for items, and thus they cannot be applied to item cold-start scenarios. However, the spring up of new items in tag recommendation scenarios is common which is a problem that needs to be solved. For our proposed MA-CVAE, by modeling item auxiliary information with deep generative models and constraining the generation of implicit feedback using different auxiliary embeddings, the new items could be recommended using item multi-auxiliary information. Here, the results between content-based baselines and our proposed MA-CVAE are listed in Table \ref{TAB:result_with_new_items}.

Results show that Bi-GRU+Att could make good recommendations on citeulike datasets whose vocabulary sizes are smaller, however, it cannot perform well on the ml-20m dataset with a straightforward multi-class classification setting. ITAG could achieve good results on ml-20m due to the modeling of tag correlations with GRU layers, while it performs badly on citeulike datasets. These two methods cannot perform well on all datasets due to specifically designed modules for content extractions.
However, our proposed method performs much better than content-based methods on three datasets which illustrates the effectiveness of our method to recommend new items using item multi-auxiliary embeddings as surrogates for the item latent embeddings. This is achieved by
constraining the closeness of the item latent embeddings and multi-auxiliary embeddings by MSE losses and additional generation losses. Moreover, through our ablation study, we can observe that different item auxiliary information all play an important role in tag recommendations for new items.

\newcommand{\mysize}{0.29\textwidth}
\begin{figure*}[]
\centering
\subfigure[ml-20m]{
\!\!\!\!\!\!\!\!\!\!\!\! \includegraphics[width=\mysize]{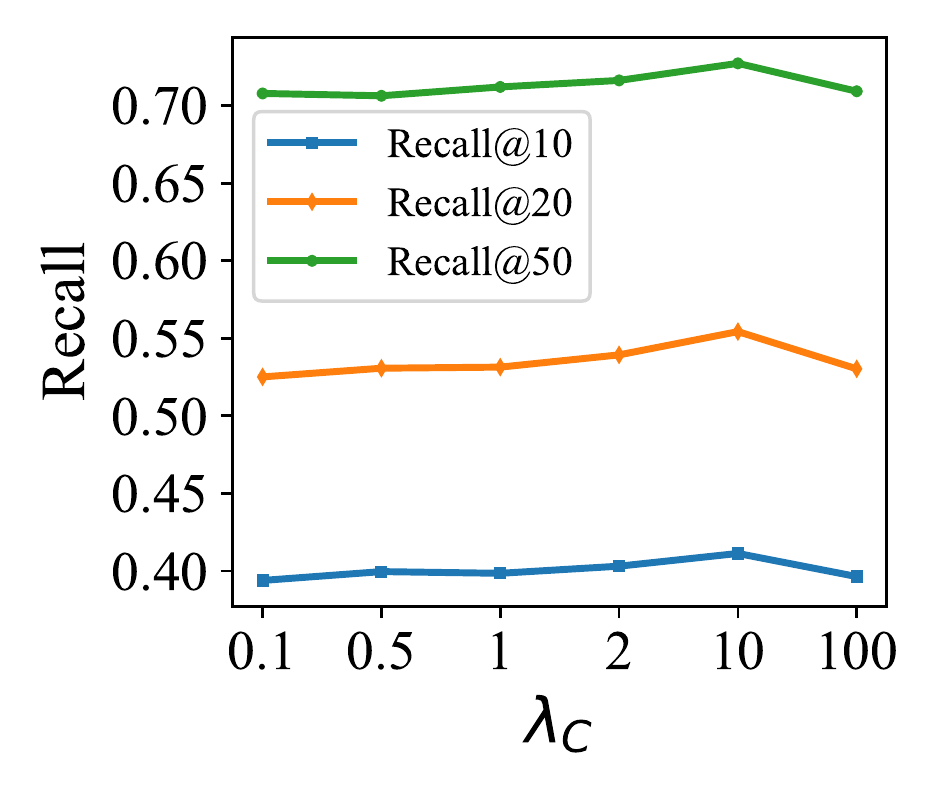}
}
\quad
\subfigure[citeulike-a]{
\!\!\!\!\!\!\!\!\!\!\!\! \includegraphics[width=\mysize]{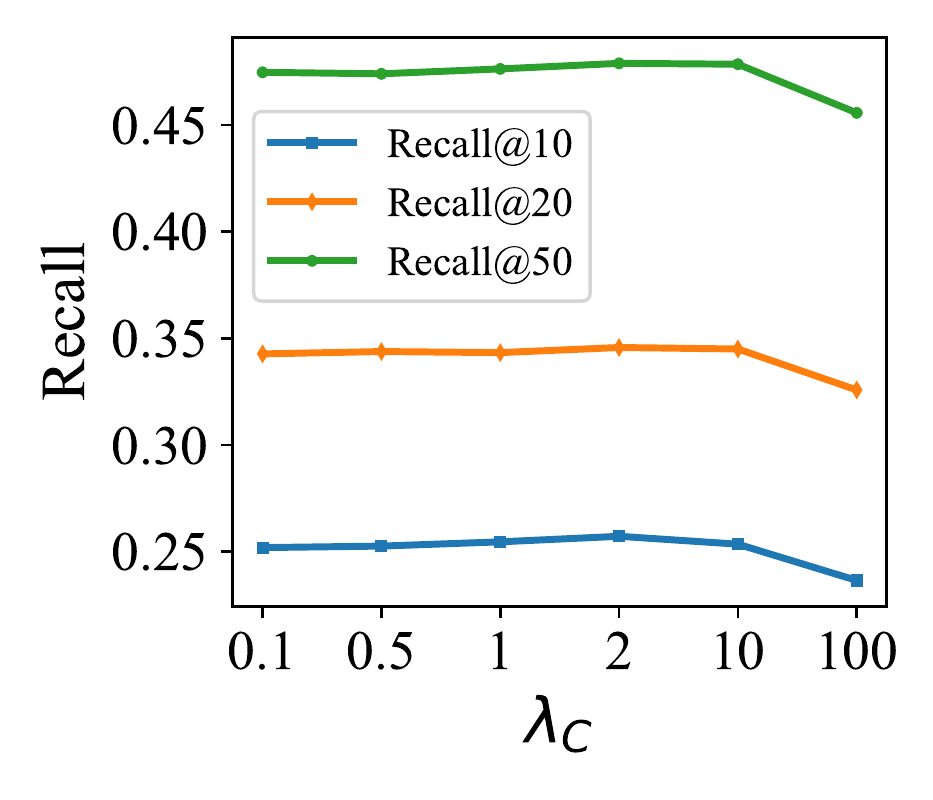}
}
\quad
\subfigure[citeulike-t]{
\!\!\!\!\!\!\!\!\!\!\!\! \includegraphics[width=\mysize]{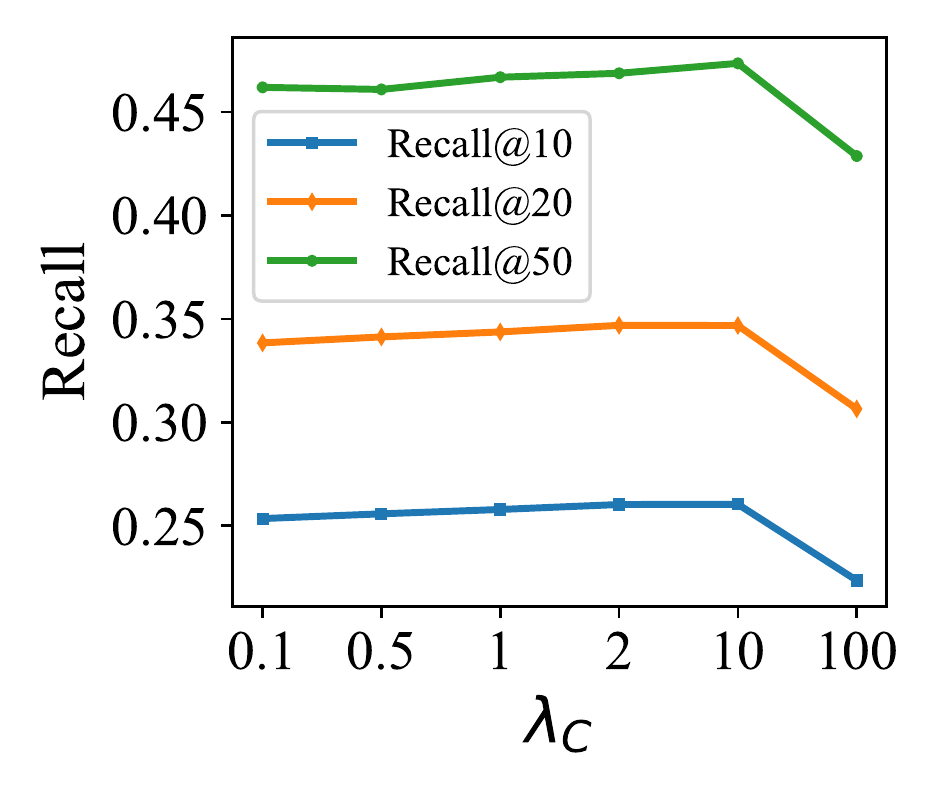}
}
\quad

\subfigure[ml-20m]{
\!\!\!\!\!\!\!\!\!\!\!\! \includegraphics[width=\mysize]{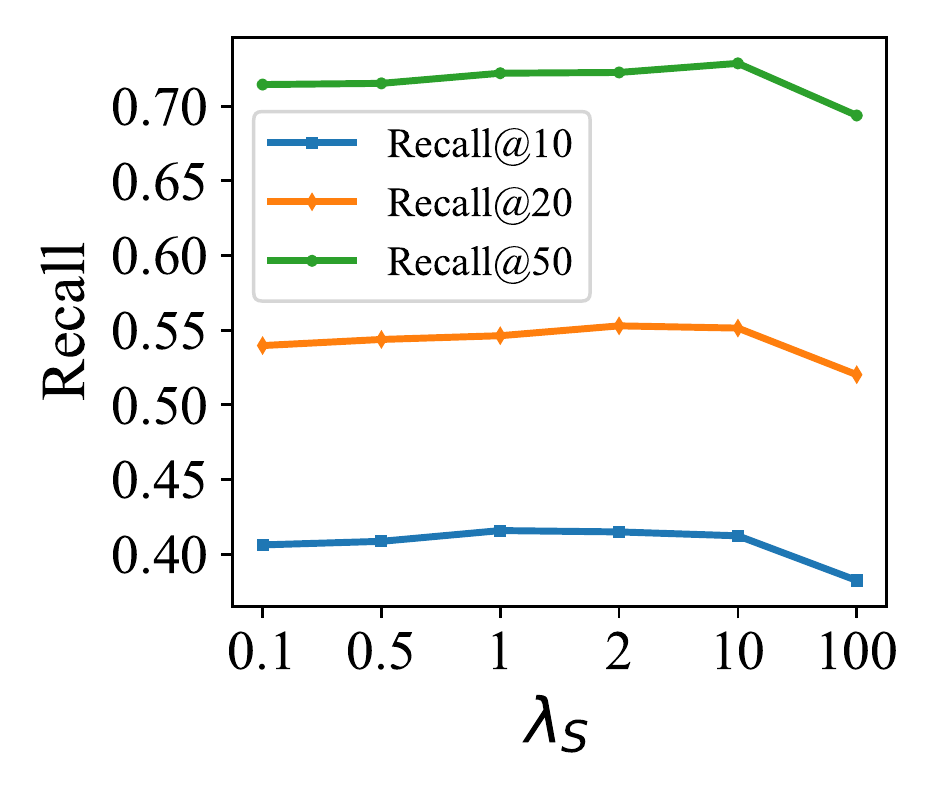}
}
\quad
\subfigure[citeulike-a]{
\!\!\!\!\!\!\!\!\!\!\!\! \includegraphics[width=\mysize]{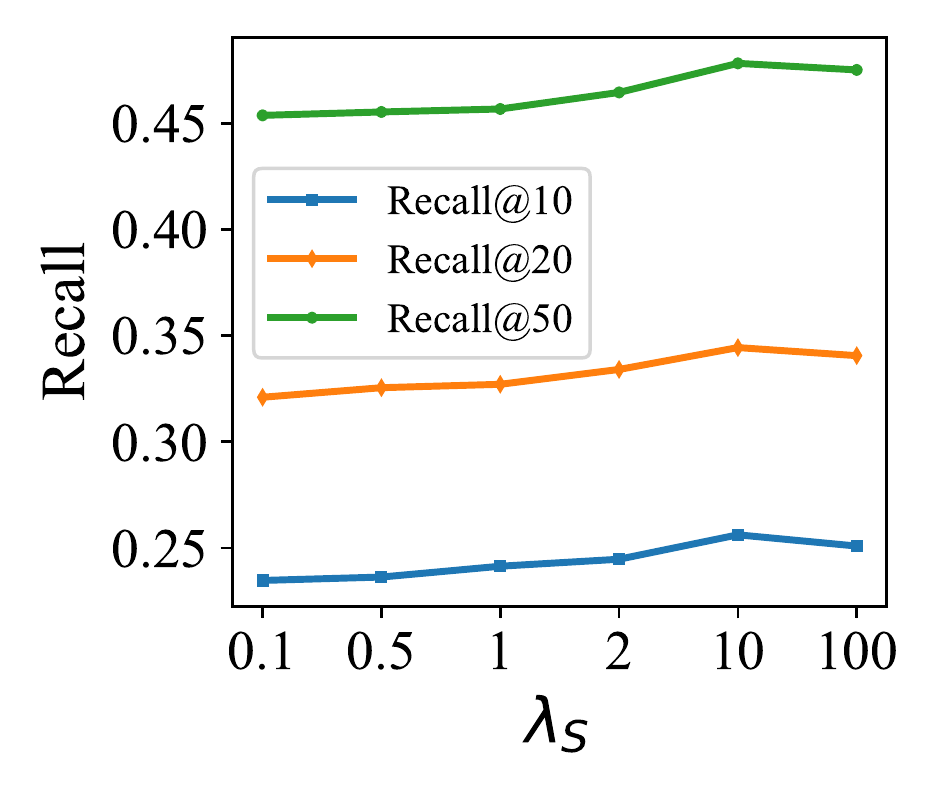}
}
\quad
\subfigure[citeulike-t]{
\!\!\!\!\!\!\!\!\!\!\!\! \includegraphics[width=\mysize]{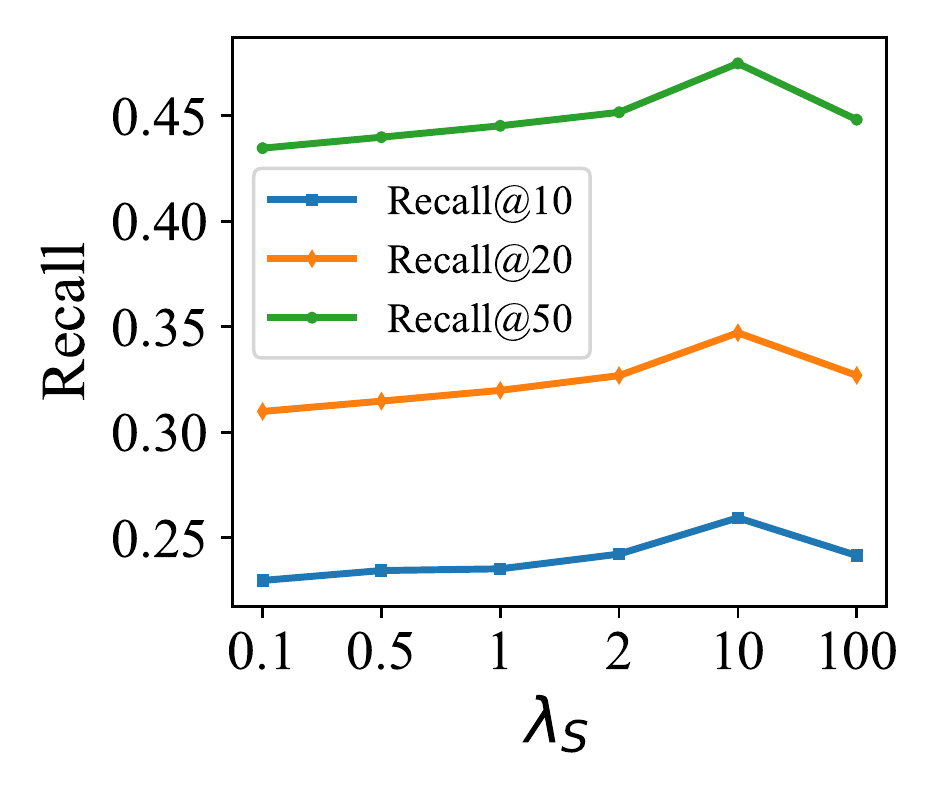}
}
\caption{Sensitivity to parameters. (a), (d) The effect of $\lambda_C$ and $\lambda_S$ on ml-20m.
(b), (e) The effect of $\lambda_C$ and $\lambda_S$ on citeulike-a. (c), (f) The effect of $\lambda_C$ and $\lambda_S$ on citeulike-t.}
\label{fig:lambda}
\end{figure*}

\subsection{Further Discussion on MA-CVAE}

\subsubsection{Sensitive to Parameters}

The parameters $\lambda_C$ and $\lambda_S$ influence the balance between the utilization of content and social graph of item multiple auxiliary information. We set $\lambda_C  \in \{0.1, 0.5, 1, 2, 10, 1000 \}$ when fixing $\lambda_S$ to be 10, and set $\lambda_S  \in \{0.1, 0.5, 1, 2, 10, 100 \}$ when fixing $\lambda_C$ to be  10. 
From Figure \ref{fig:lambda}, we have the following observations: 1) For both $\lambda_C$ and $\lambda_S$, the performance increases first and begins to decrease at some point, which demonstrates that when fixing the utilization of one auxiliary information, further exploiting another auxiliary information could improve the performance by injecting more information. However, if one kind of auxiliary information overweights the multiple information, the performance decreases much because the other information could not be utilized well and the facilitation effect between each information is reduced. 2) For citeulike datasets, the impact of performance by varying $\lambda_S$ is larger than varying $\lambda_C$, and vice-versa on ml-20m. It demonstrates that different content and social information have different importance degrees on different datasets, and therefore, it is useful to choose appropriate $\lambda_C$ and $\lambda_S$.

\subsubsection{Interpretability}
\begin{table*}[]

\renewcommand\arraystretch{1.5}

\resizebox{\textwidth}{!}{

\begin{tabular}{l|l|l}

\hline
\multirow{3}{*}{\textbf{Article   I}}           & Title                  & Evidence for dynamically organized modularity in the yeast protein-protein interaction network.                                                                  \\ \cline{2-3} 
                                       & Pred tags              & ppi, topology, pin, protein\_interaction, bio, interactome, modules, interaction-network, genetic, biological\_netwroks, graph      \\ \cline{2-3} 
                                       & True tags              & network\_analysis, protein\_interaction, pin, graph, genome, node, interactome, biological\_networks, topology,  interaction-network \\ \hline
\multirow{2}{*}{\#topology}            & Hit C                  & A link between the potential scale-free \textcolor{red}{topology} of interactome networks and genetic robustness seems to exist                                                   \\ \cline{2-3} 
                                       & {Hit S} & Ordinarily, the connection \textcolor{red}{topology} is assumed to be either completely regular or completely random                                                                                    \\ \hline
\multirow{2}{*}{\#interactome}         & Hit C                  & A link between the potential scale-free topology of \textcolor{red}{interactome} networks and genetic robustness seems to exist                                                   \\ \cline{2-3} 
                                       & Hit S                  & and hence have substantially expanded our knowledge on the protein interaction space or \textcolor{red}{interactome} of the yeast                                                 \\ \hline
\multirow{2}{*}{\#interaction-network} & Hit C                  & In apparently scale-free protein-protein \textcolor{red}{interaction networks}, or 'interactome' networks, most proteins interact with few partners                               \\ \cline{2-3} 
                                       & Hit S                  & \textcolor{red}{Interaction networks} are of central importance in postgenomic molecular biology                                                                                  \\ \hline
\multirow{3}{*}{\#graph}               & Hit C                  & None                                                                                                                                                             \\ \cline{2-3} 
                                       & \multirow{2}{*}{Hit S} & We develop a search algorithm for topological motifs called \textcolor{red}{graph} alignment                                                                                      \\ \cline{3-3} 
                                       &                        & We introduce a \textcolor{red}{graph} generating model aimed at representing the evolution of protein interaction networks                                                                                                                      \\ \hline \hline
\multirow{3}{*}{\textbf{Article II}}            & Title                  & Situated Learning: Legitimate Peripheral Participation (Learning in Doing: Social, Cognitive and Computational Perspectives)                                     \\ \cline{2-3} 
                                      & Pred tags              & learning, cultural\_studies, discourse, practice, e-learning, fieldprelim, technology,  situated, equity, digital-literacy             \\ \cline{2-3} 
                                      & True tags              & learning, cultural\_studies, information\_behavior, bibtex-import, novice, equity, digital-literacy, collaboration, practice          \\ \hline
\multirow{2}{*}{\#learning}            & Hit C                  & push forward the notion of situated \textcolor{red}{learning}--that \textcolor{red}{learning} is fundamentally a social process and not solely in the learner's head                              \\ \cline{2-3} 
                                      & Hit S                  & its very first stage is bootstrapped in a social \textcolor{red}{learning} process under the strong influence of culture.                 \\ \hline
\multirow{2}{*}{\#practice}            & Hit C                  & moving toward full participation in the sociocultural \textcolor{red}{practices} of a community                                                                                  \\ \cline{2-3} 
                                      & Hit S                  & its social organization, and the details of its implementation in actual \textcolor{red}{practice} aboard large ships                                                            \\ \hline
\multirow{3}{*}{\#cultural\_studies}   & Hit C                  & None                                                                                                                                                             \\ \cline{2-3} 
                                      & \multirow{2}{*}{Hit S} & From the chief scientist of Xerox Corporation and a research specialist in \textcolor{red}{cultural studies} at \{UC-Berkeley\} comes a treatise                                  \\ \cline{3-3} 
                                      &                        & This truly interdisciplinary text bridges art history, film, media, and \textcolor{red}{cultural studies}                                                                         \\ \hline
\multirow{3}{*}{\#cognition}           & Hit C                  & None                                                                                                                                                             \\ \cline{2-3} 
                                      & \multirow{2}{*}{Hit S} & We think the theory of distributed \textcolor{red}{cognition} has a special role to play in understanding interactions                                                            \\ \cline{3-3} 
                                      &                        & In this article we propose distributed \textcolor{red}{cognition} as a new foundation for human-computer interaction                                                            \\ \hline \hline
\multirow{3}{*}{\textbf{Article III}}           & Title                  & How to infer gene networks from expression profiles                                                                                                              \\ \cline{2-3} 
                                       & Pred tags              & gene, network, sysbio, gene\_expression, genetic, systems, causality, biology, bayesian,  statistics, regulatory\_network   \\ \cline{2-3} 
                                       & True tags              & dragan, primer, grant, systems, bayesian, sysbio, biological\_networks, gene\_expression,  regulatory\_networks       \\ \hline
\multirow{2}{*}{\#gene\_expression}    & Hit C                  & \textcolor{red}{Gene expression} data from microarrays are typically used for this purpose.                                                                                       \\ \cline{2-3} 
                                       & Hit S                  & Large-scale \textcolor{red}{gene expression} profiling generates data sets that are rich in observed features but poor in numbers of observations                                \\ \hline
\multirow{3}{*}{\#bayesian}            & Hit C                  & None                                                                                                                                                             \\ \cline{2-3} 
                                       & \multirow{2}{*}{Hit S} & This framework builds on the use of \textcolor{red}{Bayesian} networks for representing statistical dependencies. A \textcolor{red}{Bayesian} network is a                        \\ \cline{3-3} 
                                       &                        & We start by showing how \textcolor{red}{Bayesian} networks can describe interactions between genes                                                                                \\ \hline
\multirow{3}{*}{\#systems}             & Hit C                  & None                                                                                                                                                             \\ \cline{2-3} 
                                       & \multirow{2}{*}{Hit S} & from such measurements, gene/protein interactions and key biological features of cellular \textcolor{red}{systems}                                                               \\ \cline{3-3} 
                                       &                        & genetic control networks12 and many other self-organizing \textcolor{red}{systems}                                                                                                   \\ \hline \hline
\multirow{3}{*}{\textbf{Article IV}}            & Title                  & Analysing biological pathways in genome-wide association studies                                                                                                 \\ \cline{2-3} 
                                       & Pred tags              & genetics, gwas, genomics, association, methods, review, bioinformatics, statistics, gene, analysis,  genetic, pathway\_analysis                  \\ \cline{2-3} 
                                       & True tags              & gwas, methods, biology, disease, mining, networks, genomics, statistics, genomics\_analysis, pathway-analysis, genetic              \\ \hline
\multirow{3}{*}{\#gwas}                & Hit C                  & None                                                                                                                                                             \\ \cline{2-3} 
                                       & \multirow{2}{*}{Hit S} & Genome-wide association studies (\textcolor{red}{GWAS}) have rapidly become a standard method for disease gene discovery        \\ \cline{3-3} 
                                       &                        & This review is written from the viewpoint that findings from the \textcolor{red}{GWAS} provide preliminary genetic information that                                               \\ \hline
\multirow{3}{*}{\#biology}             & Hit C                  & None                                                                                                                                                             \\ \cline{2-3} 
                                       & \multirow{2}{*}{Hit S} & facilitates the usage and the analysis of biological networks in standard systems \textcolor{red}{biology} formats (SBML, SBGN, BioPAX)                                          \\ \cline{3-3} 
                                       &                        & The molecular \textcolor{red}{biology} revolution led to an intense focus on the study of interactions between DNA, RNA and protein biosynthesis                                  \\ \hline
\multirow{3}{*}{\#statistics}          & Hit C                  & None                                                                                                                                                             \\ \cline{2-3} 
                                       & \multirow{2}{*}{Hit S} & We propose number of extensions to GSEA, including the use of different \textcolor{red}{statistics} to describe the association between          \\ \cline{3-3} 
                                       &                        & We review and discuss three analytic methods to combine preliminary GWAS \textcolor{red}{statistics} to identify genes, alleles  \\ \hline
\multirow{2}{*}{\#genetic}             & Hit C                  & the power to uncover the relatively small effect sizes conferred by most \textcolor{red}{genetic} variants                                                    \\ \cline{2-3} 
                                       & Hit S                  & when used in conjunction with large databases of protein-protein, protein-DNA, and \textcolor{red}{genetic} interactions that                \\ \hline
\end{tabular}

}
\caption{Example articles with recommended tags.}
\label{tab:vualize}
\end{table*}

In order to further intuitively analyze the performance of our proposed model, we show several visualization results of MA-CVAE in this section. These results are chosen from the test set of the citeulike-a data set, where the items are articles. We use “true tags” to indicate the tags that the user actually assigns to the article, “pred tags” to indicate the tags recommended to the user, “Hit C” to indicate the partial content of the article itself, and “Hit S” to indicate the partial content of the adjacency articles in the item social graph.
Among them, the first two items are existing items in the training set, and the last one is a new item.

From Table \ref{tab:vualize}, we can see that our proposed method can recommend suitable tags for items with a good hit ratio of our predicted tags on true tags. Moreover, MA-CVAE could recommend some appropriate tags which are not included in the true tags, such as ``ppi (protein-protein-interaction)" for Article I. 
Furthermore, by observing ``Hit C" and ``Hit S" which represent the partial content of the hit sentence of the content itself and the content of its adjacent neighborhood items, we can figure out that using item content only cannot always hit the right tag. However, through properly modeling the content of the social network, the content of neighborhoods could be aggregated which contributes to making much better recommendations. For example, the tag ``graph" needs to sufficiently model item social network information to be properly recommended, which is performed by a variational graph auto-encoder model in our proposed MA-CVAE. 
On the other hand, for Article IV which is a new item, we can see that our method can still recommend suitable tags with a high hit ratio, and the item social network information (\textit{i.e.,} co-author) performs a significant role in recommendations. Since the intrinsic links in the item social graph naturally represent the proximity between items, and similar tags are well recommended for totally new items.

\section{CONCLUSIONS}
\label{section5}

In this paper, we seamlessly integrate the collaborative information and item multi-auxiliary information by defining a probabilistic generative process, where the item latent embedding that contain collaborative information and multiple auxiliary embeddings are tightly coupled by constraining their closeness with MSE losses. Specifically, implicit item-by-tag feedback, item content and social graph information are well modeled by specific-designed variational auto-encoders respectively. 
Moreover, new items could be recommended by constraining the generation of implicit interactions through the item decoder by multiple auxiliary embeddings in the training phase, as well as designing an inductive variational graph auto-encoder to enable the inference of new items. Extensive experiments on real-world datasets, MovieLens and citeulike, successfully demonstrate the effectiveness of our proposed models on both existing and new items.

In future work, we would like to extend MA-CVAE to disentangled settings, where personality and matching are considered as two aspects of tag recommendations. Specifically, personality represents taggers' personal preference among tags, while matching denotes the representative and properly-matched tags of items. The disentanglement for the two aspects should be adaptively learned for users since different demands that whether personality is more important or not is different among users.

\bibliography{ms}

\end{document}